\documentstyle[prb,preprint,aps]{revtex}
\font\ee=msbm10 scaled \magstep1
\font\ee=msbm10 scaled \magstep1

\begin{document}
\title
{\bf \Large Finite Dimensional Schwinger Basis, Deformed Symmetries, 
Wigner Function, and an Algebraic Approach to Quantum Phase}   
\author{T. Hakio\u{g}lu}
\address{Physics Department, 
Bilkent University, 06533 Ankara, Turkey} 
\maketitle
\begin{abstract}
Schwinger's finite (D) dimensional periodic Hilbert Space representations 
are studied on the toroidal lattice with specific emphasis on the deformed  
oscillator subalgebras and the generalized representations of the Wigner 
function. These subalgebras are shown to be admissible endowed with the
non-negative norm of Hilbert space vectors. Hence, they provide the desired 
canonical basis for the algebraic formulation of the quantum phase problem.    
Certain equivalence classes in the space of labels are identified within each  
subalgebra, and connections with area preserving canonical transformations 
are studied. The generalised representations of the Wigner function are 
examined in the finite dimensional cyclic Schwinger basis. These 
representations are shown to conform to all fundamental conditions of the 
generalized phase space Wigner distribution. As a specific application of the 
Schwinger basis, the number-phase unitary operator pair is studied and, based 
on the admissibility of the underlying q-oscillator subalgebra, an algebraic 
approach to the unitary quantum phase operator is established. Connections 
with the Susskind-Glogower-Carruthers-Nieto phase operator formalism as well 
as the standard action-angle Wigner function formalisms are examined in the 
infinite period limit. The concept of continuously shifted Fock basis is   
introduced to facilitate the Fock space representations of the Wigner function. 
\end{abstract}
\newpage

\section{Introduction}
Recently, finite dimensional quantum group symmetries 
find increasing physical 
applications in condensed matter systems. The Landau problem is known 
to have the $\omega_{\infty}$ symmetry\cite{1} in the algebra satisfied by 
magnetic translation operators.\cite{2} An $sl_{q}(2)$ realisation of the 
same problem has also been recently studied.\cite{3} 
The finite dimensional representations of  
these algebras are parameterised by a discrete set of labels on a    
two-dimensional toroidal lattice 
$\mbox{\ee Z}_{D}\times \mbox{\ee Z}_{D}$. 
The action of the group 
elements on the Hilbert space vectors is cyclic, the periodicity of which is  
determined by the dimension of the corresponding algebra. In the Landau 
problem the periodicity is directly connected with the degeneracy of the 
Landau levels in the ground state.\cite{3}   
In a more general framework, similar algebraic structures have been  
examined a long time ago by Schwinger in the unitary cyclic representations   
of the Weyl-Heisenberg (WH) algebra.\cite{4} Recently Floratos  
has examined\cite{5} the WH algebra parameterised by the labeling 
vectors on the toroidal lattice in terms of $D^{2}-1$ unitary traceless 
generators as a convenient representation of $su(D)$. More generally, the  
elements of the discrete and finite dimensional WH algebra are generators 
of the area preserving 
diffeomorphisms on $\mbox{\ee Z}_{D}\times \mbox{\ee Z}_{D}$ which are 
known to respect the Fairlie-Fletcher-Zachos sine algebra.\cite{6}   
The infinite dimensional extension of this is the group of infinitesimal 
area-preserving diffeomorphisms which has been examined by Arnold in the   
theory of phase space formulation of classical Hamiltonian flow.\cite{7}  
With the connection to area preserving diffeomorphisms, the finite 
dimensional WH algebra defines, in the quantum domain, the 
set of linear canonical transformations on the discrete canonical phase 
space pair, the generalised coordinate and the momentum. This has been 
observed 
as an emerging {\it presymplectic} structure preserving the discrete phase 
space area of which the connection with the classical symplectic structure  
is established in the continuous limit as the dimension of the algebra is  
extended to infinity.\cite{8} 

A more general frame for unitary cyclic reps of finite dimensional algebras,  
of which a special case is the  
WH algebra, is Schwinger's finite, special-unitary-canonical 
basis.\cite{4} Schwinger's approach proves to be a    
generalised realisation for the group of discrete area
preserving transformations on the 2-torus. This basis has   
been used indirectly in various applications 
to physics, particularly to condensed matter\cite{9,10,11} 
and field theory related problems\cite{12}    
such as the discretised versions of the Chern-Simons theory\cite{9,10},  
the dynamics of Bloch electrons in two dimensions 
interacting with a constant uniform magnetic field\cite{10},  
the quantum Hall effect\cite{11} etc. Most of these applications refer to 
the discrete WH algebra although the results can be equally valid using the 
more general Schwinger basis which will be discussed first in Sec.II.   

In this work we will follow a different route than the standart 
applications above and demonstrate that the Schwinger basis also 
provides an algebraic approach to the canonical 
phase space formulation of the celebrated quantum phase problem. 
As the first step in this route, the subalgebraic realisations of 
Schwinger's unitary operator basis 
will be constructed in Sec.III.A,B with a particular emphasis on the  
realisations in terms of the q-oscillator. It will be shown that 
in finite dimensions, these deformed oscillator realisations naturally lead
to an admissible (i.e. non-negative) cyclic spectrum by
the natural emergence of a positive Casimir operator. The net effect of the 
positive Casimir operator  
is to shift the spectrum to the admissible ranges, viz., a strong condition
on the non-negative norm of the vectors in the Hilbert space. The crucial
role played by the admissible cyclic representations in the 
canonical formulation of the quantum phase problem will be examined.  
In order to complete the picture, we also briefly discuss  
the well-known $u_{q}(sl(2))$ subalgebraic realisations of the Schwinger 
operator basis. 

The equivalence classes and  
their connection to canonical transformations on the discrete lattice 
will be discussed in Sec.III.C. Sec.IV is devoted to the application 
of the Schwinger basis in the Wigner-Kirkwood construction of the Wigner 
function. It will be shown that this construction complies with all 
fundamental properties of the Wigner function. In Sec.V we explore the   
applications of Schwinger's formalism in the    
unitary finite dimensional number-phase operator basis. In this context, 
we elaborate more on the q-oscillator subalgebraic realisations of 
Sec.III.B. In the first three subsections A,B,C of 
Sec.VI we examine the infinite dimensional limit of the number-phase basis, 
the q-oscillator subalgebra and the Wigner function respectively. 
There, it will be shown that as the 
dimension of the unitary number-phase operator algebra is extended to 
infinity, the conventional phase operator formalism of Susskind-Glogower- 
Carruthers-Nieto is recovered. We consider this as the first step to 
establish the desired unification  
of the quantum phase problem with the canonical action-angle quantum phase 
space formalism. The admissible q-oscillator subalgebra is also investigated 
in the $D \to \infty$ limit and shown to have a linear equidistant spectrum 
accompanied by a typical spectral singularity at $D=\infty$. This singular   
behaviour is examined using Fujikawa's index theorem.   

The representations of the Wigner function in the phase and number 
eigenbases are investigated in the finite and infinite Hilbert space 
dimensions. Within this unification scheme the concept of continuously 
shifted finite dimensional Fock basis is introduced in Sec.V.D.  
It is suggested that this concept facilitates the 
formulation of the Wigner function in the Fock eigenbasis. 
In the following, we start our discussion with a short study of  
Schwinger's cyclic unitary operator basis.    

\section{Finite dimensional Schwinger operator basis}
In this formulation\cite{4,8,13}, one 
considers a unitary cyclic operator $\hat{\cal U}$ acting on a finite  
dimensional Hilbert space ${\cal H}_{D}$  
spanned by a set of orthonormal basis vectors 
$\{\,\vert u\rangle_{k}\,\}_{k=0, \dots,(D-1)}$ 
with the cyclic property  
$\hat{\cal U}^{D}=\hat{\mbox{\ee I}}$ as   
\begin{equation}
\hat{\cal U}\,\vert u\rangle_{k}=
\vert u\rangle_{k+1}~,\qquad  
\vert u\rangle_{k+D}=\vert u\rangle_{k}~,\qquad 
_{k}\langle u\vert u \rangle_{k^{\prime}}=\delta_{k,k^{\prime}}. 
\label{1.1}
\end{equation}
In the $\{\vert u\rangle_{k}\}$ basis, $\hat{\cal U}$ is represented by 
\begin{equation}
\hat{\cal U}=
\sum_{k=0}^{D-1}\,\vert u \rangle_{k+1}~_{k}\langle u \vert~. 
\label{1.2}
\end{equation}
The action of $\hat{\cal U}$ corresponds to a    
{\it rotation} in ${\cal H}_{D}$. 
The axis of rotation is along the direction in ${\cal H}_{D}$     
given by the vector $\vert v\rangle_{\ell}~$ of which the direction remains  
invariant under the action of $\hat{\cal U}$ as
\begin{equation}
\hat{\cal U}\,\vert v \rangle_{\ell}=e^{i\gamma_{0} \ell}\,
\vert v \rangle_{\ell}~,\qquad 
\qquad 
\vert v\rangle_{\ell}=\frac{1}{\sqrt{D}}\,
\sum_{k=0}^{D-1}\,v_{k}^{(\ell)}\,\vert u\rangle_{k}~,\qquad 
0 \le \ell \le D-1
\label{1.3}
\end{equation}
where $v_{k}^{(\ell)}=e^{-i\gamma_{0}\,k\,\ell}$ and $\gamma_{0}=2\pi/D$. 
On the other hand it was shown by Schwinger that the new set  
$\{\,\vert v\rangle_{\ell}~\}_{\ell=0,\dots,(D-1)}$ also
forms an orthonormal set of vectors i.e. 
$_\ell\langle v\vert v\rangle_{{\ell}^{\prime}}=\delta_{\ell,\ell^{\prime}}$, 
for which one can define a second unitary operator $\hat{\cal V}$ such that  
$\hat{\cal V}^{D}=\mbox{\ee I}$ and     
\begin{equation}
\begin{array}{rl}
\hat{\cal V}\,\vert v\rangle_{\ell}=&
\vert v \rangle_{\ell+1}~,\qquad \qquad
\vert v\rangle_{\ell+D}=\vert v \rangle_{\ell}  \\
\hat{\cal V}\,\vert u\rangle_{k}=&e^{-i\gamma_{0} k}\,\vert u\rangle_{k}
\end{array}
\label{1.4}
\end{equation}
where $k \in \mbox{\ee Z}$ and $0 \le k \le (D-1)$. 
The basis vectors $\{\,\vert u\rangle_{k}\,\}_{k=0,\dots,(D-1)}$ 
and $\{\,\vert v\rangle_{\ell}\,\}_{\ell=0,\dots,(D-1)}$ define two 
equivalent and {\it conjugate} representations in the sense that 
the representation in the 
$\{\,\vert u\rangle_{k}\,\}_{k=0,\dots,(D-1)}$ basis in Eq.\,(3) is 
complemented by 
\begin{equation}
\vert u\rangle_{k}=\frac{1}{\sqrt{D}}\,
\sum_{\ell=0}^{D-1}\,u_{\ell}^{(k)}\,\vert v\rangle_{\ell}
\qquad u_{\ell}^{(k)}=e^{i\,\gamma_{0}\,\ell\,k}=
v_{k}^{(\ell)^{*}}~.
\label{1.5}
\end{equation}
The corresponding operators $\hat{\cal U}$ and $\hat{\cal V}$ satisfy 
\begin{equation}
\begin{array}{rl}
\hat{\cal U}^{m_{1}}\,\hat{\cal V}^{m_{2}}=e^{i\,\gamma_{0}\,m_{1}\,m_{2}}
\hat{\cal V}^{m_{2}}\,\hat{\cal U}^{m_{1}} \qquad 
\\
\hat{\cal U}^{m_{1}+D}=\hat{\cal U}^{m_{1}}~\qquad {\rm and} \qquad 
\hat{\cal V}^{m_{2}+D}=\hat{\cal V}^{m_{2}}~.
\end{array}
\label{1.6}
\end{equation}
An operator $\Psi$, of which the projection in the $\vert u\rangle_{k}$ 
representation is $\Psi(u_{k})$,   
is given in the $\vert v\rangle_{\ell}$ representation as 
${\tilde \Psi}(v_{\ell})$. These two conjugate representations are then 
connected by  
\begin{equation}
\Psi(u_{k})=\frac{1}{\sqrt{D}}\,\sum_{\ell=0}^{D-1}\,
_{k}\langle u\vert v\rangle_{\ell}\,{\tilde \Psi}(v_{\ell})~\qquad~, \qquad 
{\rm where} \qquad _{k}\langle u\vert v\rangle_{\ell}=
e^{-i\gamma_{0}\,\ell\,k}~.
\label{1.7}
\end{equation} 
In analogy with the elements of the discrete Wigner-Kirkwood basis,
\cite{14,15} we now define the operator\cite{4,13}   
\begin{equation}
\hat{S}_{\vec m} \equiv e^{-i\,\gamma_{0}\,m_{1}\,m_{2}/2}\,
\hat{\cal U}^{m_{1}}\,\hat{\cal V}^{m_{2}}=
e^{i\,\gamma_{0}\,m_{1}\,m_{2}/2}\,
\hat{\cal V}^{m_{2}}\,\hat{\cal U}^{m_{1}}~,
\label{1.8}
\end{equation}
where ${\vec m}=(m_{1},m_{2})$. 
We now represent the transformation in Eq's\,(\ref{1.3}) and (\ref{1.5})
between $\{\vert v \rangle_{\ell}\}_{0 \le \ell \le (D-1)}$ and 
$\{\vert u\rangle_{k}\}_{0 \le k \le (D-1)}$ bases using the unitary Fourier  
operator $\hat{\cal F}$ defined as,\cite{16}
$\{\,\vert v\rangle_{k}\,\} \equiv 
\hat{\cal F}\,\{\,\vert u \rangle_{k}\,\}$~, and 
$\{\,\vert u\rangle_{k}\,\} \equiv 
\hat{\cal F}^{-1}\,\{\,\vert v\rangle_{k}\,\}$~, 
where $\hat{\cal F}^{\dagger}=\hat{\cal F}^{-1}$. Then,  
\begin{equation}
\begin{array}{rl}
&\vert u \rangle_{k} \stackrel{\hat{\cal F}}{\longrightarrow} 
\vert v \rangle_{k} \stackrel{\hat{\cal F}}{\longrightarrow} 
\vert u\rangle_{-k} \stackrel{\hat{\cal F}}{\longrightarrow} 
\vert v\rangle_{-k} \stackrel{\hat{\cal F}}{\longrightarrow} 
\vert u \rangle_{k}~, \\
&\vert u \rangle_{k} \stackrel{\hat{\cal F}^{-1}}{\longrightarrow}
\vert v \rangle_{-k} \stackrel{\hat{\cal F}^{-1}}{\longrightarrow} 
\vert u\rangle_{-k} \stackrel{\hat{\cal F}^{-1}}{\longrightarrow} 
\vert v\rangle_{k} \stackrel{\hat{\cal F}^{-1}}{\longrightarrow} 
\vert u \rangle_{k}~. 
\end{array}
\label{1.9}
\end{equation} 
The Eq's\,(\ref{1.9}) produce a Fourier automorphism at the operator level 
as 
\begin{equation}
\begin{array}{rl}
&\hat{\cal U} \stackrel{\hat{\cal F}}{\longrightarrow} 
\hat{\cal V} \stackrel{\hat{\cal F}}{\longrightarrow} 
\hat{\cal U}^{-1} \stackrel{\hat{\cal F}}{\longrightarrow} 
\hat{\cal V}^{-1} \stackrel{\hat{\cal F}}{\longrightarrow} 
\hat{\cal U} ~, \\
&\hat{\cal U} \stackrel{\hat{\cal F}^{-1}}{\longrightarrow}
\hat{\cal V}^{-1} \stackrel{\hat{\cal F}^{-1}}{\longrightarrow} 
\hat{\cal U}^{-1} \stackrel{\hat{\cal F}^{-1}}{\longrightarrow} 
\hat{\cal V} \stackrel{\hat{\cal F}^{-1}}{\longrightarrow} 
\hat{\cal U}~. 
\end{array}
\label{1.10}
\end{equation}
Next, we define a transformation $R_{\pi/2}$ in the space of the 
lattice vector 
${\vec m}$ such that $R_{\pi/2}: (m_1,m_2) \to (-m_2,m_1)$. 
It is possible to show that 
\begin{equation}
\hat{\cal F}\,\hat{S}_{\vec m}\,\hat{\cal F}^{-1}=
\hat{S}_{R_{\pi/2}:{\vec m}}~,
\qquad \hat{\cal F}^{4}=\mbox{\ee I}~,\qquad {\rm and} \qquad 
R_{\pi/2}^{4}=\mbox{\ee I}~.
\label{1.11}
\end{equation}
Eq's\,(\ref{1.11}) imply that Eq.\,(\ref{1.8}) is invariant under 
simultaneous operations  
of $\hat{\cal F}$ and $R_{\pi/2}^{-1}$. $\hat{S}_{\vec m}$ has the 
properties 
\begin{equation}
\begin{array}{rll}
\hat{S}_{\vec m}^{\dagger}=&\hat{S}_{-{\vec m}} \\
Tr\Bigl\{\,\hat{S}_{\vec m}\,\Bigr\}=&D\,\delta_{{\vec m},{\vec 0}} \\
\hat{S}_{\vec m}\,\hat{S}_{\vec m^{\prime}}=&
e^{i\,\gamma_{0}\,{\vec m}\times{\vec m^{\prime}}/2}\,
\hat{S}_{{\vec m}+{\vec m^{\prime}}} \\
(\hat{S}_{\vec m}\,\hat{S}_{{\vec m}^{\prime}})\,
\hat{S}_{{\vec m}^{\prime \prime}}=&
\hat{S}_{\vec m}\,(\hat{S}_{{\vec m}^{\prime}}\,
\hat{S}_{{\vec m}^{\prime \prime}}) \qquad &{\rm (associativity)}\\
\hat{S}_{\vec 0}=&\mbox{\ee I} &{\rm (unit~element)}\\
\hat{S}_{\vec m} \hat{S}_{-\vec m}=&\mbox{\ee I} &{\rm (inverse)}
\end{array}
\label{1.12}
\end{equation}
where ${\vec m}\times{\vec m^{\prime}}\equiv (m_{1} m_{2}^{\prime}-
m_{2} m_{1}^{\prime})$. Using Eq's\,(\ref{1.8}) and (\ref{1.12}) it is 
possible to see that 
\begin{equation}
(\hat{S}_{\vec m}\,)^{D}=\hat{S}_{D{\vec m}}=
\hat{S}_{-D{\vec m}}=(-1)^{D m_1m_2}\,\mbox{\ee I}
\label{1.13}
\end{equation}
where $\hat{S}_{D {\vec m}}$ commutes with all elements 
$\hat{S}_{{\vec m}^{\prime}}$ for all ${\vec m}$ and ${\vec m}^{\prime}$. 
With the 
associativity condition in Eq's\,(\ref{1.12}) satisfied, the unitary 
Schwinger 
operator basis $\hat{S}_{\vec m}$ defines a discrete projective 
representation of the Heisenberg algebra parameterised by the 
{\it discrete phase space} vector ${\vec m}$ in 
$\mbox{\ee Z}_{D} \times \mbox{\ee Z}_{D}$.  
Excluding ${\vec m}={\vec 0}$ and if $D$ is a prime number, 
the elements of the basis $\hat{\cal U}^{m_1}\,\hat{\cal V}^{m_2}$  
form a complete set of $D^{2}-1$ unitary traceless matrices providing 
an irreducible representation for $su(D)$. If $D$ is not a prime, 
then the prime decomposition of $D$ as 
$D=D_{1} D_{2} \dots D_{i} \dots$, as shown in Ref.s\,[4,13], 
permits the study of a physical system with a number of quantum degrees of 
freedom with each degree 
of freedom expressed in terms of an independent Schwinger basis with the 
cyclic property determined by the particular prime factor $D_{j}$.        
In what follows, we will assume that $D$ is a prime number representing 
 a single degree of freedom. Exceptional cases will be independently  
mentioned as needed.  

The eigenspace of $\hat{S}_{\vec m}$ is spanned by the eigenvectors 
$\vert {\vec m}, r\rangle_{\{0 \le r \le (D-1)\}}$ with eigenvalues 
$\lambda_{r}({\vec m})$. Using Eq's\,(\ref{1.1}) and (\ref{1.4}) we expand 
the eigenvectors $\vert {\vec m}, r\rangle$ where $0 \le r \le (D-1)$ in, 
for instance, the $\vert u \rangle_{k}$ basis with coefficients  
$e^{(r)}_{k}({\vec m}) \equiv ~_{k}\langle u \vert {\vec m}, r\rangle$. 
From this definition and Eq.\,(\ref{1.1}) 
it is clear that the coefficients are periodic, 
i.e., $e^{(r)}_{k}({\vec m})=e^{(r)}_{k+D}({\vec m})$.  
The coefficients and eigenvalues are then determined by the recursion  
\begin{equation}
\lambda_{r}({\vec m})\,e^{(r)}_{k}({\vec m})=
e^{-i\beta_{k}( m_1, m_2)}\,e^{(r)}_{k-m_1}({\vec m}) ~,\quad {\rm where}  
\quad \beta_{k}(m_1,m_2)=\gamma_{0}m_2(2\,k-m_1)/2~. 
\label{1.14}
\end{equation} 
which yields 
\begin{equation}
\lambda_{r}({\vec m})=e^{i\pi\,m_1\,m_2}\,e^{-i\frac{2\pi\,r}{D}}
~,\qquad  
e^{(r)}_{k}({\vec m})=\Bigl\{\,\prod_{n=0}^{M-1}\,\lambda_{r}({\vec m})\,
e^{-i\beta_{k-n\,m_1}( m_1, m_2)}\,\Bigr\}\,e^{(r)}_{k+1}({\vec m})
\label{1.15}
\end{equation}
where $M=[mod(D)+1]/m_1$ and $M \in \mbox{\ee Z}$. In deriving 
Eq.\,(\ref{1.15}) from (\ref{1.14}) we used the 
periodicity property $e_{k}^{(r)}({\vec m})=e_{k+D}^{(r)}({\vec m})$. 
It should be noted that the diagonal representations 
$\vert {\vec m}, r\rangle$ of $\hat{S}_{\vec m}$ in the 
$\vert u \rangle_{k}$  
and $\vert v\rangle_{k}$ bases are equivalent and consistent 
with Eq's\,(\ref{1.9}) and (\ref{1.10}) only for the case when $D$ 
is a prime number. We will come back to Eq.\,(\ref{1.15}) when we examine  
the q-oscillator subalgebraic realisations of the Schwinger basis in 
Sec. III.B. We now turn to the subalgebraic structure of the Schwinger 
basis.  

\section{The deformed subalgebraic structure}
It is well-known that the $\hat{S}_{\vec m}$ basis  
has an explicit deformed algebraic structure. Defining the operators  
$\hat{D}_{\vec m}=D/2\pi\,\hat{S}_{\vec m}$, the commutator    
\begin{equation}
\bigl[\hat{D}_{\vec m},\hat{D}_{\vec n}]=i~\frac{2}{\gamma_{0}}\,
\sin(\frac{\gamma_{0}}{2}{\vec m}\times {\vec n})~\,
\hat{D}_{{\vec m}+{\vec n}}~,
\label{2.1}
\end{equation}
describes the Fairlie-Fletcher-Zachos sine algebra.\cite{6} 
The generators of the algebra $\hat{J}_{\vec m}$ can be represented by  
the Weyl matrices\cite{17} 
\begin{equation} 
g=\pmatrix{1& 0 & 0 & \dots & 0 \cr
0& \omega & 0 & \dots & 0 \cr
0 & 0 & \omega^{2} & \dots & 0\cr
\vdots & \vdots & \vdots & \ddots & 0 \cr
0 & 0 & 0 & \dots & \omega^{D-1}\cr} 
\qquad 
h=\pmatrix{0 & 1 & 0 & \dots & 0\cr
0 & 0 & 1 & \dots & 0 \cr
\vdots & \vdots & \vdots & \ddots & 0\cr
0 & 0 & 0 & \dots & 1 \cr
 1 & 0 & 0 & \dots & 0\cr}
\label{2.2}
\end{equation}
with 
$\hat{J}_{\vec m}=\omega^{m_{1}m_{2}/2}\,g^{m_1}\,h^{m_2}$ satisfying 
$h\,g=\omega\,g\,h$, $g^{D}=h^{D}=\mbox{\ee I}$, with    
$\omega^{D}=1$ and $\omega=e^{i\,\gamma_{0}}$. With these at hand,  
it is possible to verify that  
$[\hat{J}_{\vec m},\hat{J}_{\vec n}]=
i\,2/\gamma_{0}\,\sin(\gamma_{0}/2\,{\vec m}\times 
{\vec n})\,\hat{J}_{{\vec m}+{\vec n}}$. 

The deformed $u_{q}(sl(2))$ subalgebraic realisations of the sine algebra
have been under extensive investigation
recently, based on the magnetic translation operator basis.\cite{9,10,11,12}
 In the following we will present a brief account of 
this symmetry in the more general Schwinger basis.  

\subsection{The $u_{q}(sl(2))$ subalgebraic realisation}
We define the operators $\hat{A}$ and $\hat{A}^{\dagger}$ as
\begin{equation}
\hat{A}\equiv d \hat{S}_{\vec m}+d^{\prime} \hat{S}_{{\vec m}^{\prime}}~,
\qquad
\hat{A}^{\dagger} \equiv d^{*} \hat{S}_{-\vec m}+ d^{\prime^{*}}
\hat{S}_{-{\vec m}^{\prime}}
\label{2.3}
\end{equation}
where $d$ and $d^{\prime}$ satisfy  
\begin{equation}
d\,d^{{\prime}^{*}}=d^{*}\,d^{\prime}=- 
(p^{1/2}-p^{-1/2})^{-2}~,\qquad \qquad 
p=e^{-i\gamma_{0}\,{\vec m}\times {\vec m}^{\prime}}~.  
\label{2.4}
\end{equation}
We find that
\begin{equation}
\begin{array}{rll}
\hat{A}\,\hat{S}_{{\vec m}-{\vec m}^{\prime}}=
&p\,\hat{S}_{{\vec m}-{\vec m}^{\prime}}\,
\hat{A}~,\qquad \qquad 
&\hat{A}\,\hat{S}_{{\vec m}-{\vec m}^{\prime}}^{\dagger}=
p^{-1}\,\hat{S}_{{\vec m}-{\vec m}^{\prime}}^{\dagger}\,\hat{A} 
\qquad \\
\hat{S}_{{\vec m}-{\vec m}^{\prime}}=&s_{p}\,
p^{\hat{J}_{3}}~,\qquad {\rm where} 
&{\hat A}\,\hat{J}_{3} \equiv (\hat{J}_{3}+1)\,\hat{A}~.  
\end{array}
\label{2.5}
\end{equation}
and $s_{p}=e^{-i\pi{\vec m}\times {\vec m}^{\prime}}=p^{D/2}$,  
such that Eq.\,(\ref{1.13}) holds. It is also possible to realise in 
Eq's\,(\ref{2.5}) 
that $\hat{S}_{\vec m}\,\hat{S}_{-{\vec m}^{\prime}}=
\tilde{s}_{p}\,p^{\hat{J}_{3}}$ such that $\tilde{s}_{p}=
e^{-i\pi\gamma_{0}(D-1){\vec m}\times {\vec m}^{\prime}}=p^{(D-1)/2}$. 

For both cases, a direct calculation yields    
\begin{equation}
\bigl[\hat{A},\hat{A}^{\dagger}]=-\frac{p^{\hat{J}_{3}+D/2}-
p^{-\hat{J}_{3}-D/2}}
{p^{1/2}-p^{-1/2}}\equiv -[\hat{J}_{3}+\frac{D}{2}]~,
\label{2.6}
\end{equation}
which together with Eq's\,(\ref{2.5}) implies an $u_{p^{1/2}}(sl(2))$ 
symmetry 
defined by the elements $\hat{A}, \hat{A}^{\dagger}, \hat{J}_{3}$.   
The Casimir operator for this subalgebra is given by
\begin{equation}
C_{p}=\hat{A}^{\dagger} \hat{A}+
[\frac{1}{2}(\hat{J}_{3}+\frac{D}{2}-\frac{1}{2})]^{2}=
\hat{A} \hat{A}^{\dagger}+
[\frac{1}{2}(\hat{J}_{3}+\frac{D}{2}+\frac{1}{2})]^{2}
\label{2.7}
\end{equation}
where $[x]$ is formally given in Eq.\,(\ref{2.6}). The Hilbert space 
is spanned by 
the vectors $\vert j,j_{3}\rangle$ where $\hat{J}_{3}\,\vert j,j_{3}\rangle
=j_{3}\,\vert j,j_{3}\rangle$ with $-j \le j_{3} \le j$. If the lowest 
weight representations exist such that     
 $\hat{A}\,\vert j,-j \rangle\equiv0$, then $j$ is determined by the 
value of the Casimir operator as 
$\hat{C}_{p}\,\vert j,-j\rangle=
[1/2(D/2-1/2-j)]^{2}\,\vert j,-j\rangle$. The lowest    
weight representations are obtained by successive operations 
of $\hat{A}^{\dagger}$ on the state $\vert j, -j\rangle$. 
These representations 
are $D$ dimensional for the particular case $j=(D-1)/2$ such that 
$\hat{A}^{\dagger}\,\vert 
j,-j+(D-1)\,\rangle=\hat{A}^{\dagger}\,\vert j,j\rangle=0$, 
where the highest and lowest weight reps coincide. In this case  
the representations are cyclic with period $D$.    
For this case, the Casimir operator vanishes.    
We close this section by referring to the extensive 
applications of the $u_{q}(sl(2))$ symmetry, for instance in 
Ref's.\,[9--12] and move on to another subalgebraic realisation of the 
Schwinger basis. 

\subsection{The spectrum shifted admissible q-oscillator realisation}
Let's now consider the $\hat{A}$ and $\hat{A}^{\dagger}$ operators in 
Eq.\,(\ref{2.3}) where  
$d$ and $d^{\prime}$ are constant to be redetermined for the q-oscillator 
realisation. Using Eq's\,(\ref{1.12}) we construct the q-commutator   
\begin{equation}
\begin{array}{rl}
\hat{A}\hat{A}^{\dagger}-q\hat{A}^{\dagger}
\hat{A}=&(|d|^{2}+|d^{\prime}|^{2})\,(1-q) \\
&+d\,d^{\prime^{*}}\,(e^{-i\gamma_{0}{\vec m}\times {\vec m}^{\prime}}-q)\,
\hat{S}_{-{\vec m}^{\prime}}\,\hat{S}_{\vec m}
+d^{\prime} d^{*}\,(e^{i\gamma_{0}{\vec m}\times {\vec m}^{\prime}}-q)\,
\hat{S}_{-\vec m} \, \hat{S}_{{\vec m}^{\prime}}
\end{array}
\label{2.8}
\end{equation} 
where ${\vec m}\times {\vec m}^{\prime} \ne (mod D)$. 
Here, $\vert q \vert=1$ and is otherwise arbitrary at this level. 
Eq's\,(\ref{2.8}) can be written as
\begin{equation}
\hat{A}\hat{A}^{\dagger}-q\hat{A}^{\dagger}\hat{A}=
(|d|^{2}+|d^{\prime}|^{2})\,(1-q)+\hat{Q}~, \qquad
\qquad {\rm for} \qquad q=
e^{\pm\,i\gamma_{0}\,{\vec m}\times {\vec m}^{\prime}}
\label{2.9}
\end{equation}
where  
\begin{equation}
\hat{Q}=\cases{d\,d^{\prime^{*}}\,(q^{-1}-q)\,\hat{S}_{-{\vec m}^{\prime}}
\,\hat{S}_{\vec m}~, & if ~~ 
$q=e^{i\gamma_{0}{\vec m}\times {\vec m}^{\prime}}$ \cr  
d^{\prime} d^{*}\,(q^{-1}-q)\, \hat{S}_{-{\vec m}}\,\hat{S}_{
{\vec m}^{\prime}}~, & if ~~ 
$q=e^{-i\gamma_{0}{\vec m}\times {\vec m}^{\prime}}$~. \cr}  
\label{2.10}
\end{equation}
It can be shown that 
\begin{equation}
\hat{A} \hat{Q}= q^{-1}\,\hat{Q} \hat{A}~, 
\qquad q=e^{\pm i \gamma_{0}{\vec m}\times {\vec m}^{\prime}}
\label{2.11}
\end{equation}
which implies that a generalised number operator $\hat{N}$ can be defined
in such a way that
$\hat{A}\hat{N} \equiv (\hat{N}+1)\hat{A}$ and $\hat{Q}=c_{q}\,q^{-\hat{N}}$,
where $c_{q}$ is a proportionality constant whose value 
depends on the choice of $d$ and $d^{\prime}$. Eq.\,(\ref{2.11}) implies that 
$\hat{A}^{D}, \hat{A}^{\dagger^{D}}$ commute with all elements of the 
algebra. Since the cases $q$ and $q^{-1}$ give rise to identical 
results as far as the algebra is concerned, we only examine the case
$q=e^{-i\gamma_{0}{\vec m}\times {\vec m}^{\prime}}$. 
In order to determine $c_{q}$ we first make the choice  
\begin{equation}
d\,d^{\prime^{*}}=\frac{1}{q^{-1}-q}=
\frac{1}{2i\,\sin(\gamma_{0}{\vec m}\times {\vec m}^{\prime})}~, \qquad
{\rm hence} \qquad \vert d \vert \, \vert d^{\prime}\vert=
\frac{1}{2\,\vert \sin(\gamma_{0}{\vec m}\times {\vec m}^{\prime})\vert}~.
\label{2.12}
\end{equation}
The constants $d, d^{\prime}$ are also undetermined up to a constant overall
phase factor. Choosing their magnitudes symmetrically we can determine
the real positive shift constant $C$ as
\begin{equation}
C=\vert d\vert^{2}+\vert d^{\prime}\vert^{2}=
\frac{1}{\vert \sin(\gamma_{0}{\vec m}\times {\vec m}^{\prime})\vert}~.  
\label{2.13}
\end{equation}
The first one in Eq's\,(\ref{2.10}) leads to the same result
in (\ref{2.13}). From Eq's\,(25) we have 
$\hat{Q}^{D}=c_{q}^{D}\,q^{-D\,\hat{N}}$. 
Then, making use of $q^{D} \equiv 1$ and Eq's\,(\ref{1.12}) and 
(\ref{1.13}), we find that 
$c_{q}=e^{i\gamma_{0}(D-1){\vec m}\times {\vec m}^{\prime}/2}=q^{-(D-1)/2}$.  
It can be seen that the net effect of the pure phase $c_{q}$ is to shift 
the spectrum of $\hat{N}$ by an overall constant $(D-1)/2$. Hence,  
$\hat{Q}=q^{-\hat{N}-(D-1)/2}$.   

With the generalised number operator as defined below Eq.\,(\ref{2.11}), 
we have   
\begin{equation}
\begin{array}{rl}
&\hat{A}\hat{A}^{\dagger}-q\hat{A}^{\dagger}\hat{A}=C\,(1-q)+
q^{-\hat{N}-(D-1)/2} \\
\hat{A}\,\hat{N}&=(\hat{N}+1)\,\hat{A}~, \qquad
\hat{A}^{\dagger}\,\hat{N}=(\hat{N}-1)\,\hat{A}^{\dagger}~. 
\end{array}
\label{2.14}
\end{equation}
Eq's\,(\ref{2.14}) describe the q-oscillator algebra with its 
spectrum shifted by the positive constant $C$ as  
\begin{equation}
\hat{A}^{\dagger}\hat{A}=C+[\hat{N}]~,\qquad {\rm where} \qquad 
[\hat{N}]=\frac{q^{\hat{N}+(D-1)/2}-q^{-\hat{N}-(D-1)/2}}
{q-q^{-1}}
\label{2.15}
\end{equation}
where $0 \le \Vert\hat{A}^{\dagger} \hat{A} \Vert$ as required, and,  
$C$ is identified with the central invariant,  
which plays a crucial role in the existence of the admissible 
cyclic reps of the q-oscillator algebra endowed with a positive 
spectrum.\cite{18,19}  
In Eq's\,(\ref{2.14}), the existence of the lowest (highest) weight vectors 
such that $\hat{A}\,\vert n_{0}\rangle=
\hat{A}^{\dagger}\,\vert n_{0}+D-1\rangle=0$ crucially depends on the 
specific values of $D$ and ${\vec m}\times {\vec m}^{\prime}$.   
The condition for the existence of such $\vert n_{0}\rangle$ is 
given by $C=-[n_{0}]$. For $C$ as given by (\ref{2.13}), it can be checked 
in Eq.\,(\ref{2.15}) that 
this condition is violated for $D$ being an odd number. If $D$ is 
an even number, such reps are permitted for 
${\vec m}\times {\vec m}^{\prime}=D/2 (mod D)$, however in that case 
they are not irreducible.  
For $D$ being a prime other than two, the situation is the same as when 
$D$ is odd. We now examine how the q-oscillator algebra generators $\hat{A}, 
\hat{A}^{\dagger}$ and $\hat{N}$ act in the eigenspace of $\hat{S}_{\vec m}$ 
operators. We first observe that if 
$\vert {\vec m}-{\vec m}^{\prime} , r\rangle$ is an 
eigenstate of $\hat{S}_{{\vec m}-{\vec m}^{\prime}}$ with eigenvalue 
$\lambda_{r}({\vec m}-{\vec m}^{\prime})$ for $0 \le r \le D-1$,  
\begin{equation}
\begin{array}{rl}
\hat{S}_{{\vec m}-{\vec m}^{\prime}}\,
\vert {\vec m}-{\vec m}^{\prime},r\rangle\equiv &
\lambda_{r}({\vec m},{\vec m}^{\prime})\,
\vert {\vec m}-{\vec m}^{\prime} , r\rangle \\
\hat{S}_{\vec m}\,\vert {\vec m}-{\vec m}^{\prime} , r\rangle=&
g_{r}({\vec m},{\vec m}^{\prime})\,\vert {\vec m}-{\vec m}^{\prime},
r-{\vec m}\times {\vec m}^{\prime} \rangle \\
\hat{S}_{{\vec m}^{\prime}}\,\vert {\vec m}-{\vec m}^{\prime} , r\rangle=&
f_{r}({\vec m},{\vec m}^{\prime})\,
\vert {\vec m}-{\vec m}^{\prime} , 
r-{\vec m}\times {\vec m}^{\prime} \rangle \\
\end{array}
\label{2.16}
\end{equation}
where the second and third equations can be deduced from Eq's\,(\ref{1.12}). 
In the second and third equations, 
$g_{r}({\vec m},{\vec m}^{\prime})$ and $f_{r}({\vec m},{\vec m}^{\prime})$ 
are pure phase factors to be determined. Using Eq's\,(\ref{2.3}) we 
compare the action of $\hat{A}$ in the q-oscillator eigenbasis 
$\vert n \rangle$ and in the eigenbasis $\vert {\vec m},r\rangle$ as 
\begin{equation}
\begin{array}{rl}
\hat{A}\,\vert {\vec m}-{\vec m}^{\prime}, r\rangle=&(d\,g+d^{\prime}\,f)\,
\vert {\vec m}-{\vec m}^{\prime}, r-{\vec m}\times {\vec m}^{\prime} \rangle
\\
\hat{A}\,\vert n\rangle=&\sqrt{C+[n]}\,\vert n-1 \rangle
\end{array}
\label{2.17}
\end{equation} 
where it is directly implied that a unit shift in $n$ corresponds to 
a shift of $r$ in units of ${\vec m}\times {\vec m}^{\prime}$. Since 
${\vec m}\times {\vec m}^{\prime} \ne ( mod D)$ by construction, the set 
of integers $n\,{\vec m}\times {\vec m}^{\prime}$ for $0 \le n \le (D-1)$ 
is the same as $n$ itself. Then, all  
eigenvectors in the q-oscillator and the Schwinger bases are  
connected on a one-to-one basis  
with successive operations of $\hat{A}$ and $\hat{A}^{\dagger}$. 
Since the eigenbasis $\{\vert {\vec m}, r\rangle\}_{0 \le r \le (D-1)}$ 
is normalised, Eq's\,(\ref{2.17}) imply that
\begin{equation}
\vert (d\,g+d^{\prime}\,f)\vert^{2}=C+[n]~.
\label{2.18}
\end{equation}
We then apply Eq's\,(\ref{2.12}) and (\ref{2.13}) to obtain
\begin{equation}
\frac{\vert g\vert^{2}+\vert f\vert^{2}-i(g\,f^{*}-g^{*}\,f)}
{2\,\vert \sin(\gamma_{0}{\vec m}\times {\vec m}^{\prime})\vert}=
\frac{\vert 1+\sin[\gamma_{0}\,(n+(D-1)/2)\,{\vec m}\times {\vec m}^{\prime}]
\vert}{\vert \sin[\gamma_{0}\,{\vec m}\times {\vec m}^{\prime})\vert}~.
\label{2.19}
\end{equation}
Since $\vert g\vert=\vert f\vert=1$, Eq.\,(\ref{2.19}) yields  
\begin{equation}
g_{r}({\vec m},{\vec m}^{\prime})=f_{r}^{*}({\vec m},{\vec m}^{\prime})=
e^{i\frac{\gamma_{0}}{2}\,(n+(D-1)/2){\vec m}\times {\vec m}^{\prime}}~,
\label{2.20}
\end{equation}
where it can be considered that $r=n\,{\vec m}\times {\vec m}^{\prime}$. 
Comparing $\lambda_{r}({\vec m})$ in Eq's\,(\ref{1.15}) with the first 
equation in (\ref{2.16}) we find that 
$\lambda_{r}({\vec m},{\vec m}^{\prime})=
e^{i\gamma_{0}(n-D/2){\vec m}\times{\vec m}^{\prime}}$.  
Eq's\,(\ref{2.16}--\ref{2.20}) indicate that the admissible cyclic 
representations of the q-oscillator realisation for a fixed value of the 
deformation parameter $q \ne 1$ have one-to-one correspondence with the 
diagonal representations of the Schwinger basis for a fixed but arbitrary  
non-collinear vectors ${\vec m}, {\vec m}^{\prime}$. 
  
The admissible q-oscillator subalgebraic structure of the 
Schwinger basis has not been taking too much attention. 
The $su_{q}(2)$ realisations of two shifted and mutually commuting 
q-oscillators in the Schwinger boson representation has been studied by 
Fujikawa\cite{20} recently. It will be demonstrated in Sec.V that this 
particular realisation plays a crucial role in the canonical formulation 
of the quantum phase problem. 

\subsection{Equivalence Classes and Canonical Transformations on the Lattice}
In both the q-oscillator and the $u_{p^{1/2}}(sl(2))$ cases examined here,   
there are sets of equivalence classes 
$E_{{\vec m}\times {\vec m}^{\prime}}$ 
incorporating those sets of subalgebras  
parameterised by different lattice vector pairs ${\vec m}$   
and ${\vec m}^{\prime}$ such that the  
deformation parameter remains invariant under unitary transformations 
within each such class. 

Let's assume a transformation 
$R^{q}_{\vec{m}, \vec{m}^{*}; \vec{m}^{*}, \vec{m}^{*^{\prime}}}$ whose   
effective action is to map the pair ${\vec m}, {\vec m}^{\prime}$ 
into a new one ${\vec m}^{*}, {\vec m}^{\prime^{*}}$ in  
$\mbox{\ee Z}_{D}\times \mbox{\ee Z}_{D}$ as  
\begin{equation}
R^{p}_{\vec{m}, \vec{m}^{\prime}; \vec{m}^{*}, \vec{m}^{*^{\prime}}}\,
f(\vec{m}, \vec{m}^{\prime})=f(\vec{m}^{*}, \vec{m}^{*^{\prime}})
\label{2.21}
\end{equation}
such that $\vec{m}\times {\vec m}^{\prime}=
\vec{m}^{*}\times {\vec m}^{*^\prime}$, hence  
$\vec{m}, \vec{m}^{\prime};\vec{m}^{*}, \vec{m}^{*^\prime} \in 
E_{\vec{m}\times {\vec m}^{\prime}}$. Here $f$ represents an arbitrary 
function. If  
$R^{p}_{\vec{m}, \vec{m}^{\prime}; \vec{m}^{*}, \vec{m}^{*^{\prime}}}$  
is represented in Eq.\,(\ref{2.21}) by the $2\times 2$ 
integer matrix $R$, then 
the $R$ matrix satisfies 
\begin{equation}
R^{t}\,P\,R=P~, \qquad {\rm where} \qquad P=\pmatrix{0 & 1 \cr -1& 0\cr}
\label{2.22}
\end{equation}
with $\det{R}=+\,1$. Here $R^{t}$ corresponds to the ordinary transpose 
of $R$. Eq.\,(\ref{2.22}) implies that both $\vec{m}$ and 
$\vec{m}^{\prime}$ are to be transformed by the same transformation  
\begin{equation}
R\,f(\vec{m},\vec{m}^{\prime})=f(R\,\vec{m}, R\,\vec{m}^{\prime})=
f(\vec{m}^{*},\vec{m}^{*^\prime})~, 
\label{2.23}
\end{equation}
and besides the unimodularity of $R$ there is no further restriction.  
Hence, $R$ is an element of $sl(2,\mbox{\ee Z}_{D})$. 
The product ${\vec m}\times {\vec m}^{\prime}$ 
corresponds to the exact cocycle\cite{8} in the Schwinger operator basis  
which is proportional to the discrete phase space area spanned by the vectors 
${\vec m}, {\vec m}^{\prime}$. Hence, $R$ plays the role of a class of 
area-preserving canonical transformations. As a result of the projective 
character of the Schwinger basis, any unitary transformation acting on 
the basis elements preserves the phase space area hence the symplectic 
structure as described by the matrix $P$ in Eq.\,(\ref{2.22}). 
The action of $R$ on the lattice is then equivalent to the reflection    
of such unitary transformations in the operator space.  

At this point, we find it necessary to mention briefly that there are 
implications of the equivalence classes in the construction of the  
generalised coproduct $\Delta(.^{\small \otimes})$ for two deformed  
subalgebras parameterised by different lattice vectors. Let's denote by  
$\hat{X}_{1}, \hat{X}_{1}^{\dagger}, \hat{H}_{1}$ and 
$\hat{X}_{2}, \hat{X}_{2}^{\dagger}, \hat{H}_{2}$ as the generators in   
two such algebras from the same equivalence  
class. It is possible to write for their tensor product algebra a 
generalised coproduct as 
\begin{equation}
\begin{array}{rl}
\Delta(\hat{X}^{\small ^{\otimes}})=
&\hat{X}_{1} \otimes p^{\hat{H}_{2}/2}+
p^{-\hat{H}_{1}/2} \otimes \hat{X}_{2} \\
\Delta(\hat{X}^{\dagger^{\otimes}})=&\hat{X}_{1}^{\dagger} \otimes
p^{\hat{H}_{2}/2}+p^{-\hat{H}_{1}/2} \otimes \hat{X}_{2}^{\dagger} \\
\Delta(\hat{H}^{\small ^{\otimes}})=&\hat{H}_{1} \otimes \mbox{\ee I}+
\mbox{\ee I}\otimes \hat{H}_{2}~,
\end{array}
\label{2.24}
\end{equation}
where $\Delta(\hat{X}^{\small ^{\otimes}}), 
\Delta(\hat{X}^{\dagger^{\otimes}}), \Delta(\hat{H}^{\small ^{\otimes}})$ 
respect the same deformed algebra. 

Keeping their labels on the lattice explicit, we now consider all operators 
$\hat{A}_{{\vec n},{\vec n}^{\prime}}$, 
$\hat{A}_{{\vec n},{\vec n}^{\prime}}^{\dagger}$ and 
$\hat{S}_{{\vec n}-{\vec n}^{\prime}}$ on the translated lattice   
by ${\vec r}$ such that  ${\vec n}={\vec m}+{\vec r}$ and 
${\vec n}^{\prime}={\vec m}^{\prime}+{\vec r}$. 
The algebra on this translated lattice space is given by   
\begin{equation}
\begin{array}{rl}
\hat{A}_{{\vec m}+{\vec r},{\vec m}^{\prime}+{\vec r}}
\hat{S}_{{\vec m}-{\vec m}^{\prime}}=&p^{\prime}\,
\hat{S}_{{\vec m}-{\vec m}^{\prime}}\,
\hat{A}_{{\vec m}+{\vec r},{\vec m}^{\prime}+{\vec r}} ~, \qquad 
\hat{A}^{\dagger}_{{\vec m}+{\vec r},{\vec m}^{\prime}+{\vec r}} 
\hat{S}_{{\vec m}-{\vec m}^{\prime}}=p^{\prime^{-1}}\,
\hat{S}_{{\vec m}-{\vec m}^{\prime}}\,
\hat{A}^{\dagger}_{{\vec m}+{\vec r},{\vec m}^{\prime}+{\vec r}} \\
\hat{S}_{{\vec m}-{\vec m}^{\prime}}=&s_{{p}^{\prime}}\,
p^{\prime^{\hat{J}_{3}}}~,\qquad \qquad 
p^{\prime}=p\,e^{i\,\gamma_{0}\,\delta \alpha},\qquad 
\qquad \delta \alpha={\vec r}\times ({\vec m}-{\vec m}^{\prime})  
\end{array}
\label{2.25}
\end{equation}
Here, $p$ is given in Eq.\,(\ref{2.4}). The Eq's\,(\ref{2.25}) define the 
elements of $u_{p^{\prime^{1/2}}}(sl(2))$ with a different 
deformation parameter 
$p^{\prime}$. It is clear that translations on the lattice   
are not in the class of area-preserving transformations defined  
above, and they cannot be realised by any unitary transformation on the
Schwinger basis. Such transformations act as a bridge between the    
two projective representations characterised by 
two different cocycles. In our formalism here, this effectively corresponds 
to transforming the elements of those subalgebras belonging to one 
equivalence 
class $E_{{\vec m}\times {\vec m}^{\prime}}$ into those of the other one    
$E_{{\vec n}\times {\vec n}^{\prime}}$. In the example of 
Eq.\,(\ref{2.25}), these two subalgebras are $u_{p^{1/2}}(sl(2))$ 
and $u_{p^{\prime^{1/2}}}(sl(2))$ with deformations 
$p$ and $p^{\prime}=p\,e^{i\gamma_{0}\delta \alpha}$ respectively.   

We now shift our attention to a more general structure of linear canonical 
transformations implicitly generated by $R$ on the lattice.
The similarity tranformation induced by the Fourier operator $\hat{\cal F}$ 
in Eq.s\,(\ref{1.9}--\ref{1.11}) has been shown in Sec.\,II to effectively 
generate the simplest example of canonical transformations,   
i.e. a $\pi/2$ rotation on  
$\mbox{\ee Z}_{D}\times \mbox{\ee Z}_{D}$. Let's now seek   
general canonical transformations on the lattice generated by an 
operator $\hat{\cal G}$ such that
\begin{equation}
\hat{\cal G}\,\hat{S}_{\vec m}\,\hat{\cal G}^{-1}=
\hat{S}_{R:{\vec m}}=\hat{S}_{{\vec m}^{\prime}}~,\qquad {\rm where} \qquad 
R\,=\pmatrix{s_1&t_1 \cr s_2&t_2\cr} \in sl(2,\mbox{\ee Z}_{D})
\label{2.26}
\end{equation} 
with ${\vec s}\times {\vec t}=\det\,R=1$, where ${\vec s}=(s_1,s_2)$ and 
${\vec t}=(t_1,t_2)$ are two vectors on 
$\mbox{\ee Z}_{D}\times \mbox{\ee Z}_{D}$. 
Such a transformation $\hat{\cal G}$ can be  
given more explicitly in the $\hat{\cal U}, \hat{\cal V}$ basis by
\begin{equation}
\begin{array}{rl}
\hat{\cal G}\,\hat{U}\,\hat{\cal G}^{-1}=\hat{S}_{\vec s}~,\qquad \qquad   
\hat{\cal G}\,\hat{V}\,\hat{\cal G}^{-1}=\hat{S}_{\vec t} \\
\hat{\cal U} \, \stackrel{\hat{\cal G}}{\longrightarrow} \, \hat{S}_{\vec s} 
\, \stackrel{\hat{\cal G}}{\longrightarrow} \, 
\hat{S}_{s_1{\vec s}+s_2{\vec t}} \stackrel{\hat{\cal G}}{\longrightarrow} 
...\\ 
\hat{\cal V} \, \stackrel{\hat{\cal G}}{\longrightarrow} \, \hat{S}_{\vec t} 
\, \stackrel{\hat{\cal G}}{\longrightarrow} \, 
\hat{S}_{t_1{\vec s}+t_2{\vec t}} \stackrel{\hat{\cal G}}{\longrightarrow} 
...
\end{array}
\label{2.27}
\end{equation}
Using Eq's\,(\ref{2.27}) and the results in Sec.\,II, the action of the 
$\hat{\cal G}$ 
operator on the basis vectors $\{\vert u\rangle_{k}\}_{0 \le k \le (D-1)}$ 
and $\{\vert v\rangle_{k}\}_{0 \le k \le (D-1)}$ can be found to be  
\begin{equation}
\hat{\cal G}\,\vert v \rangle_{k}=\vert {\vec s}, k\rangle~,\qquad 
\hat{\cal G}\,\vert u \rangle_{k}=\vert {\vec t}, -k\rangle
\label{2.28}
\end{equation}
where, similarly to the first one of Eq's\,(\ref{2.16}), 
$\vert {\vec s}, k\rangle$ and $\vert {\vec t}, -k\rangle$ are the 
eigenvectors of $\hat{S}_{\vec s}$ and $\hat{S}_{\vec t}$ with 
eigenvalue indices $k$ and $-k$ respectively.  
Hence $\hat{\cal G}$ converts the vectors in the eigenbasis of 
$\hat{\cal U}$ and $\hat{\cal V}$ into those in $\hat{S}_{\vec s}$ and 
$\hat{S}_{\vec t}$ respectively. The similarity transformation in 
Eq's\,(\ref{1.9}-\ref{1.11}) 
is a special case of the transformation in (\ref{2.26}) and (\ref{2.27}) 
for ${\vec s}=(0,1)$ and ${\vec t}=(-1,0)$. 

\section{Applications to the Wigner-Kirkwood Basis and the 
generalised Wigner function}
We now consider the discrete Wigner-Kirkwood operator basis\cite{13,14,15} 
$\hat{\Delta}({\vec V})$   
acting on the quantum phase space spanned by the vectors 
${\vec V}=(V_1,V_2)$. Phase space representations in   
the Wigner-Kirkwood and the Schwinger bases are connected by the dual form   
\begin{equation}
\hat{\Delta}({\vec V})=\frac{1}{D^{2}}\,
\sum_{\vec m}\,e^{-i\gamma_{0}\,({\vec m}\times {\vec V})}\,\hat{S}_{\vec m}
~,\qquad
\hat{S}_{\vec m}=\int d{\vec V}\,
e^{i\gamma_{0}\,{\vec m}\times {\vec V}}\,\hat{\Delta}({\vec V})
\label{3.1}
\end{equation}
where ${\vec m}\times {\vec V}=m_1 V_2-m_2 V_1$ and the range of the integral 
over ${\vec V}$ is the entire 2-torus. Similar constructions in the 
discrete formalism have also been made, for instance in Ref's.\,[8,13].  
The Wigner function $W_{\psi}({\vec V})$ is defined as the projection 
of $\Delta({\vec V})$ in a physical state $\vert \psi \rangle$ as  
\begin{equation}
W_{\psi}({\vec V})=\langle \psi \vert 
\hat{\Delta}({\vec V}) \vert \psi \rangle~. 
\label{3.2}
\end{equation}
Any operator $\hat{F}(\hat{\cal U}, \hat{\cal V})$ with 
$\Vert \hat{F}\Vert < \infty$ can then be associated with a classical 
 function $f({\vec V})$ as
\begin{equation}
\frac{1}{D}\,\langle \psi \vert \hat{F} \vert \psi \rangle=\int\,
d{\vec V}\,f({\vec V})\,W_{\psi}({\vec V})~,\qquad \quad 
f({\vec V})=
Tr\Bigl\{\hat{F}\,\hat{\Delta}^{\dagger}({\vec V})\,\Bigr\}~; \qquad 
d{\vec V}=dV_1\,dV_2
\label{3.3}
\end{equation}
Hereupon, the particular normalization we will use is based on 
$\int_{0}^{2\pi}\,dx\,e^{i\,x\,n}=2\pi\delta_{n,0}$ and in the continuous 
limit 
$\lim_{D\to \infty}\,\sum_{n=0}^{D-1}\,e^{i\,x\,n}=2\pi\delta(x)$. 

Let's now consider the action of $\hat{\cal F}$ in Sec.II. 
The action of the Fourier operator $\hat{\cal F}$ on the Wigner-Kirkwood 
basis can be found using Eq.\,(\ref{3.1}) to be  
$\hat{\cal F}\,\hat{\Delta}({\vec V})\,
\hat{\cal F}^{-1}=\hat{\Delta}(R_{\pi/2}^{-1}:{\vec V})$, where 
$R_{\pi/2}^{-1}: {\vec V}=(V_2,-V_1)$ with $R_{\pi/2}$ as given in 
Eq.\,(\ref{1.11}). This is one of 
the simplest non-trivial canonical transformations corresponding to the 
rotation of the vector ${\vec V}$ by $\pi/2$ on the quantum phase space. 
As an extension of the finite transformations generated 
by the operator $\hat{\cal F}$, one can find in Eq.\,(\ref{2.26}) explicit 
unitary transformations generated by $\hat{\cal G}$ 
of which the reflections on the 
quantum phase space are linear canonical ones on the quantum phase 
space observables. 

The properties of a generalised phase-space Wigner function have  
been enlisted by Hillery et al.\cite{21} under several fundamental 
conditions. Most of these conditions can be checked by employing the 
appropriate canonical transformations $\hat{\cal G}$ and the corresponding 
$R$. In the following we will check these conditions for 
Eq.\,(\ref{3.1}) using the properties of the Schwinger basis.

i) The Wigner function is real: $W_{\psi}({\vec V})=W^{*}_{\psi}({\vec V})$ 

Using the first equation in (\ref{1.12}) it can easily be proven 
that $\Delta({\vec V})$ is a self-adjoint operator. Hence, 
$W_{\psi}({\vec V})$ is real.   

ii) Integration over one phase space variable $V_{i}$ yields the marginal 
probability distribution of the physical state in the eigenbasis of the 
other variable $V_{j}$: $\int\,dV_{i}\,W_{\psi}({\vec V})=
\vert \langle V_{j}\vert \psi\rangle\vert^{2}$ where 
$\vert V_{j}\rangle=\vert v\rangle_{V_{j}}$ for $(i=1,j=2)$ and 
$\vert V_{j}\rangle=\vert u\rangle_{V_{j}}$ for $(i=2,j=1)$.  

To prove this property using Eq.\,(\ref{3.1}), perform the integral over 
$V_{i}$ to obtain $D\,\delta_{m_{j},0}$. Then express $\hat{S}_{\vec m}.\,
\delta_{m_{j},0}$ in the $\hat{\cal U}, \hat{\cal V}$ basis where only 
the $m_{j}$'th power of $\hat{\cal U}$ or $\hat{\cal V}$ appears. Write 
$\hat{\cal U}$ or $\hat{\cal V}$ raised to the power $m_{j}$ in terms of 
its eigenbasis using Eq.\,(\ref{1.1}--\ref{1.3}) or (\ref{1.4},\ref{1.5}). 
Following the $m_{i}~(i \ne j)$ 
summation, perform the summation over the eigenvector index $k$  
to obtain the proof. Note that this condition is true for any canonically 
transformed ${\vec V}=(V_{1}, V_{2})$ such that $V_{i} \to (R:{\vec V})_{i}, 
V_{j} \to (R:{\vec V})_{j}$, which can be easily done using $\hat{\cal G}$ 
and $R$ in Eq.\,(\ref{2.26}).  

iii) $W_{\psi}({\vec V})$ should be covariant under Galilean translations 
on the phase space.  

Since the phase space spanned by the vectors ${\vec m}$ is discrete,  
the translations are generated by the integer powers of 
$\hat{\cal U}$ and $\hat{\cal V}$ operators as 
$\hat{\cal U}^{n_1}\,\vert u\rangle_{k}=\vert u\rangle_{k+n_1}$ and 
$\hat{\cal V}^{n_2}\,\vert v\rangle_{k}=\vert u\rangle_{k+n_2}$.  
In the Galilean translated physical state 
$\vert \psi^{\prime}\rangle=
{\hat{\cal U}^{n_1} \choose \hat{\cal V}^{n_2}}\,\vert \psi\rangle$,  
the Wigner function is given by  
\begin{equation}
W_{\psi}^{\prime}({\vec V})=\langle \psi \vert 
{\hat{\cal U}^{-n_1} \choose \hat{\cal V}^{-n_2}}\,
\hat{\Delta}({\vec V})\,
{\hat{\cal U}^{n_1} \choose \hat{\cal V}^{n_2}}\,\vert \psi\rangle
\label{3.4}
\end{equation}
where the upper and lower cases correspond to the translations performed  
independently in either the $\vert u\rangle_{k}$ or the $\vert v\rangle_{k}$
basis. Using the properties of the $\hat{\cal U}$ and $\hat{\cal V}$ 
operators as well as Eq's\,(\ref{1.12}) it can be shown that
\begin{equation}
W_{\psi}^{\prime}({\vec V})=W_{\psi}({\vec V}^{\prime})~,\qquad {\rm where} 
\qquad {\vec V}^{\prime}={V_1+n_1,V_2 \choose V_1,V_2+n_2}~. 
\label{3.5}
\end{equation}
Hence, Eq.\,(\ref{3.1}) is covariant under Galilean translations on the 
lattice.   

iv) $W_{\psi}({\vec V})$ should be covariant under space and/or 
time inversions. 

To prove this, we assume that the time inversion is defined by 
$(m_1,m_2) \stackrel{{T}^{*}}{\longrightarrow} (m_1,-m_2)$ 
and the space inversion is given by   
$(m_1,m_2) \stackrel{{P}}{\longrightarrow} (-m_1,-m_2)$.  
The time inversion is a $\det\,T^{*}=-1$ type improper canonical 
transformation. Following a similar derivation in the time inverted,  
i.e. $\vert \psi^{\prime}\rangle=
{\cal T}^{*}\,\vert \psi\rangle$, or space inverted, i.e. 
$\vert \psi^{\prime}\rangle=\hat{\cal P}\,\vert \psi \rangle$, physical state 
$\vert \psi^{\prime}\rangle$, it is possible to see that  
\begin{equation}
\begin{array}{rl}
W_{\hat{\cal T}^{*}:\psi}({\vec V})=&W_{\psi}({\vec V}^{\prime})~, 
\qquad {\vec V}^{\prime}=(V_1,-V_2) \\ 
W_{\hat{\cal P}:\psi}({\vec V})=&W_{\psi}({\vec V}^{\prime})~, 
\qquad {\vec V}^{\prime}=(-V_1,-V_2)
\end{array}
\label{3.6}
\end{equation}
In particular we notice that the transformation corresponding to space 
inversion is 
identical to the successive operations of the Fourier operator in 
Eq.\,(\ref{1.11}) twice, viz., $\hat{\cal P}=\hat{\cal F}^{2}$.  

v) If $W_{\psi}({\vec V})$ and $W_{{\psi}^{\prime}}({\vec V})$ are two 
Wigner functions corresponding to the physical states $\vert \psi \rangle$ 
and $\vert \psi^{\prime} \rangle$ respectively, then 
\begin{equation}
\int d{\vec V}\,W_{\psi}({\vec V})\,W_{{\psi}^{\prime}}({\vec V})=
\frac{1}{D}\,
\vert \langle \psi \vert \psi^{\prime} \rangle \vert^{2}~.
\label{3.7}
\end{equation}
We present the proof starting from
\begin{equation}
\int d{\vec V}\,W_{\psi}({\vec V})\,W_{{\psi}^{\prime}}({\vec V})=
\frac{1}{D^{4}}\,\sum_{{\vec m},{\vec m}^{\prime}}\,\int\,d{\vec V}\,
e^{-i\gamma_{0}({\vec m}+{\vec m}^{\prime})\times {\vec V}}\,
\langle \psi \vert \hat{S}_{\vec m} \vert \psi\rangle\,
\langle \psi^{\prime} \vert \hat{S}_{{\vec m}^{\prime}} 
\vert \psi^{\prime}\rangle 
\label{3.8}
\end{equation}
We then express $\vert \psi \rangle$ and $\vert \psi^{\prime} \rangle$,   
for instance in the $\{\vert u\rangle_{k}\}_{0 \le k \le (D-1)}$ basis as
\begin{equation}
\vert \psi \rangle=\sum_{k}\, \psi_{k}\,\vert u\rangle_{k}~, \qquad \qquad 
\vert \psi^{\prime} \rangle=\sum_{k}\, \psi_{k}^{\prime}\,
\vert u\rangle_{k}~.
\label{3.9}
\end{equation}
The ${\vec V}$ integral yields $D^{2}\,\delta_{{\vec m},-{\vec m}^{\prime}}$.
Then, using $\hat{S}_{\vec m}\,\vert u\rangle_{k}=
e^{-i\gamma_{0}/2(2k+m_1)m_2}\,\vert u \rangle_{k+m_1}$ and performing the 
summations over $m_1,m_2$ we obtain the right hand side of Eq.\,(\ref{3.7}).   

vi) If $\hat{Y}$ and $\hat{Z}$ are two dynamical operators of $\hat{\cal U}$ 
and $\hat{\cal V}$, then
\begin{equation}
\frac{1}{D}\,Tr\Bigl\{\hat{Y}\,\hat{Z}\Bigr\}=
\int d{\vec V}\,y({\vec V})\,z({\vec V})
\label{3.10}
\end{equation}
where $y({\vec V})$ and $z({\vec V})$ are classical functions on the phase 
space corresponding to $\hat{Y}$ and $\hat{Z}$.  

The proof of this condition can be done using Eq.\,(\ref{3.3}) and  
$Tr\Bigl\{\hat{S}_{\vec m}\Bigr\}=D\,\delta_{{\vec m},{\vec 0}}$.  

We thus suggest that the realisations of the generalised Wigner-Kirkwood 
basis in 
terms of the elements of the Schwinger basis as expressed in Eq.\,(\ref{3.1}) 
satisfies all fundamental conditions to represent the Wigner function in  
a more generalised form.  

The connection between the unitary transformations 
in the Schwinger basis and canonical area preserving ones on the 
quantum phase space have been intensively studied recently. 
We refer to Ref.\,[8] for a detailed analysis of this connection.  
The Wigner function on $\mbox{\ee Z}_{D}\times \mbox{\ee Z}_{D}$ 
has been examined by Wooters\cite{22} and applications to action-angle 
case and the problems therein have been  
recently studied in detail by Bizzaro\cite{23} and Vaccaro\cite{24}. 

The discrete Wigner function we examined in this section is based on the 
particular normalization adopted in Eq's\,(\ref{3.1}), (i.e. 
the $1/D^2$ factor in the first equation). Using a different 
normalization, it is also possible to examine the case where one of the 
two (or both) continuous phase space variables 
${\vec V}=(V_1,V_2)$ is (are) replaced by the discrete ones. The former 
is more convenient in the case where the canonical variables 
correspond to the action-angle pair, whereas the latter should be used 
when the discrete phase space variables are considered on equal 
footing (i.e. canonical linear discrete coordinate and momentum\cite{13}). 
It should be noted that in Sec. V and VI   
we will use the normalization adopted for the action-angle variables and 
replace the $1/D^2$ factor in Eq's\,(\ref{3.1}) by $1/(2\pi\,D)$ in order 
to obtain the conventional action-angle Wigner function in the continuous 
limit. 

\section{Applications to the Unitary Number-Phase Basis and Connection 
with the Quantum Phase Problem}
It is known that a finite dimensional admissible cyclic algebra 
\begin{equation}
\begin{array}{rlc}
\hat{a}\,\vert n \rangle=&f(n)^{1/2}\,\vert n-1\rangle~,  \qquad & n \ne 0
\\
\hat{a}^{\dagger}\,\vert n\rangle=&f(n+1)^{1/2}\,\vert n+1 \rangle~,
\qquad & n \ne (D-1) \\
\hat{a}\,\vert 0 \rangle=&f(0)^{1/2}\,\beta\,\vert D-1\rangle~, \qquad &
 \vert \beta \vert=1; \vert D\rangle \equiv \vert 0 \rangle \\
\hat{a}^{\dagger}\,\vert D-1 \rangle=&f(D)^{1/2}\,\beta^{*}\,\vert 0
\rangle~,\\
0 \le& f(n) ~, \qquad & n \in \mbox{\ee Z}~~(mod D)
\end{array}
\label{4.1}
\end{equation}
provides a well-defined algebraic basis for the quantum phase 
operator.\cite{19,20}  
Here $\hat{a}$ and $\hat{a}^{\dagger}$ are spectrum lowering and raising 
operators and $f(n)$ is a generalised spectrum with the cyclic property 
that $f(n+D)=f(n)$. The admissibility condition is enforced by 
the last equation in (\ref{4.1}).  

The unitary phase operator $\hat{\cal E}_{\phi}$ is   
given in the generalised cyclic number basis by\cite{19,20}
\begin{equation}
\hat{\cal E}_{\phi}=
\sum_{n=0}^{D-1}\,\vert n-1\rangle
\,\langle n\vert~, \qquad \vert n+D\rangle \equiv 
\vert n\rangle~\quad {\rm for~all} \quad  n
\label{4.2}
\end{equation}
where the discrete phase eigenvalues and eigenstates are  
\begin{equation}
\hat{\cal E}_{\phi}\,\vert \phi \rangle_{\ell}=
e^{i\,\gamma_{0} \ell}\,\vert \phi \rangle_{\ell} \qquad
\vert \phi \rangle_{\ell}=\frac{1}{\sqrt{D}}\,
\sum_{n=0}^{D-1}\,e^{i\,\gamma_{0} n \ell}\,\vert n \rangle ~, 
\qquad \vert \phi\rangle_{\ell+D} \equiv \vert \phi\rangle_{\ell}
\label{4.3}
\end{equation}
with $0 \le \ell \le D-1$. 
The phase eigenbasis is orthonormal and resolves the identity as
\begin{equation}
_{\ell^{\prime}}\langle \phi \vert \phi \rangle_{\ell}=
\delta_{\ell^{\prime},\ell} \qquad {\rm and} \qquad \mbox{\ee I}=
\sum_{\ell=0}^{D-1}\,\vert \, \phi \, \rangle_{\ell} ~ 
_{\ell}\langle \, \phi \, \vert~.
\label{4.4}
\end{equation}
We now define the unitary operator 
$\hat{\cal E}_{N}=e^{-i \gamma_{0}\,\hat{N}}$  
with $\hat{N}$ describing the number operator such that  
$\hat{N}\,\vert n \rangle=n\,\vert n \rangle$. Then, 
$\hat{\cal E}_{N}=e^{-i \gamma_{0}\,\hat{N}}$ has 
\begin{equation}
\hat{\cal E}_{N} \vert \phi\rangle_{\ell}=\vert \phi \rangle_{\ell-1}~, 
\qquad  
\hat{\cal E}_{N} \vert n \rangle=e^{-i\gamma_{0}n}\,\vert n \rangle
\qquad {\rm where} \qquad 
\vert n \rangle=\frac{1}{\sqrt{D}}\,\sum_{\ell=0}^{D-1}\,
e^{-i\gamma_{0}\,n \ell}\,\vert \phi\rangle_{\ell}~.
\label{4.5}
\end{equation}
The properties of the unitary phase and number operators 
$\hat{\cal E}_{\phi}$ and $\hat{\cal E}_{N}$  
have been recently studied from this algebraic point of view.\cite{19}  
Here, in addition to these properties, they  
also establish a particular application of  
Schwinger's operator basis. Among the four equivalent choices in 
Eq's\,(\ref{1.10}), we examine the particular case 
\begin{equation}
{ \hat{\cal U} \choose \hat{\cal V}} \Rightarrow 
{\hat{\cal E}_{N} \choose \hat{\cal E}_{\phi}}~.
\label{4.6}
\end{equation}
Using this, and following (\ref{1.8}), we construct the operators 
$\hat{S}_{\vec m}$ in the number-phase basis as  
\begin{equation}
\hat{S}_{\vec m} \equiv e^{-i\,1/2\,\gamma_{0} m_{1} m_{2}}\,
\hat{\cal E}_{N}^{m_{1}}\,\hat{\cal E}_{\phi}^{m_{2}}, 
\qquad {\rm where} 
\qquad 
\hat{\cal E}_{N}^{m_{1}}\,\hat{\cal E}_{\phi}^{m_{2}}=
e^{i\gamma_{0}\,m_1 m_2}\,\hat{\cal E}_{\phi}^{m_{2}}\,
\hat{\cal E}_{N}^{m_{1}}~.
\label{4.7}
\end{equation}
All properties of the cyclic Schwinger unitary operator basis studied in 
Sec.\,II and III are satisfied in the unitary number-phase basis. 
In addition to these properties,      
a strong limitation exists on the admissibility of the representations 
in ${\cal H}_{D}$ to make the mapping in (\ref{4.6}) an acceptable one.   

The q-oscillator algebra in Sec.III.A defined by the elements  
$\hat{A}, \hat{A}^{\dagger}$, and $\hat{N}$ for a fixed ${\vec m}$ 
and ${\vec m}^{\prime}$ 
with $q=e^{\pm\,i\gamma_{0}{\vec m}\times {\vec m}^{\prime}}$ and 
$\gamma_{0}=2\pi/D$ is an admissible cyclic algebra 
which provides a natural realisation of Eq's\,(\ref{4.1}) with  
$\hat{a} \to \hat{A}$, $\hat{a}^{\dagger} \to \hat{A}^{\dagger}$, and 
$\hat{N}$. In this case,   
the admissible algebra in Eq's\,(\ref{4.1}) is given by the shifted 
q-oscillator algebra in Eq's\,(\ref{2.14}) where  
\begin{equation}
f(n) \to [n]+C=\frac{q^{n+(D-1)/2}-q^{-n-(D-1)/2}}{q-q^{-1}}+C~, 
\qquad   
C=\frac{1}{\vert \sin(\gamma_{0}{\vec m}\times {\vec m}^{\prime})
\vert} \ne 0~.  
\label{4.8}
\end{equation}
Now, let's consider a real cyclic operator $F(\hat{N})$ with 
$0 \le \Vert F(\hat{N}) \Vert$  
such that $F(\hat{N})=F(\hat{N}+D)$ of which the eigenvalues in the   
number basis $\{ \vert n \rangle\}_{0 \le n \le (D-1)}$ are given by 
$f(n)$. We consider the expansion of $F(\hat{N})$ as   
\begin{equation}
F(\hat{N})=\frac{1}{D}\,\sum_{k=0}^{D-1}~\tilde{f}_{k}~~
q^{-\hat{N}\,k}~,\qquad 
\qquad q=e^{-i\gamma_{0}\,{\vec m}\times {\vec m}^{\prime}}~. 
\label{4.9}
\end{equation} 
The sets of integers 
$\{k~{\vec m}\times {\vec m}^{\prime}; 
{\vec m}\times {\vec m}^{\prime}\ne(mod D)\}_{0 \le k \le (D-1)}$ 
and $\{k\}_{0 \le k \le (D-1)}$ are equivalent for any 
${\vec m}, {\vec m}^{\prime}$. Thus, Eq's\,(\ref{4.9}) is  
nothing but the operator Fourier expansion of $F(\hat{N})$. Using 
Eq.\,(\ref{2.10}), and the fact that ${\vec m}$ and ${\vec m}^{\prime}$ are 
not to be collinear, the 
operator $q^{-\hat{N}}$ can be realised as the third element 
$c_{q}^{-1}\,\hat{S}_{-{\vec m}^{\prime}}\,\hat{S}_{\vec m}$ of the 
q-oscillator subalgebra. Hence, Eq.\,(\ref{4.9}) can be equivalently 
written as 
\begin{equation}
F(\hat{N})=\frac{1}{D}\,\sum_{k=0}^{D-1}~\tilde{f}_{k}~~
\hat{S}_{(k{\vec m}-{\vec m}^{\prime})}~,\quad \quad 
\tilde{f}_{k}=Tr\Bigl\{ \hat{S}_{(k{\vec m}-{\vec m}^{\prime})}^{\dagger}\,
F(\hat{N})\,\Bigr\}=\sum_{n=0}^{D-1}\,e^{i\gamma_{0}\,k\,n}\,f(n)
\label{4.10}
\end{equation}
where we redefined $\tilde{f}_{k}$ as 
$\tilde{f}_{k} \to \tilde{f}_{k} c_{q}^{-1}\,q^{1/2}=
\tilde{f}_{k}\,q^{D/2}$. Since the vectors ${\vec m},{\vec m}^{\prime}$  
are fixed but indetermined, Eq.\,(\ref{4.10}) is the expansion of 
$F(\hat{N})$ 
in an arbitrary but fixed q-oscillator subalgebra based on a fixed ${\vec m}$ 
and ${\vec m}^{\prime}$ of the Schwinger basis 
with the deformation parameter 
$q=e^{-i\gamma_{0}{\vec m}\times {\vec m}^{\prime}}$. 

As a specific application of Sec.IV, and making use of the 
correspondence in (\ref{4.6}),        
we construct the Schwinger realisation of the discrete Wigner-Kirkwood 
operator basis in the number-phase space as  
\begin{equation}
\hat{\Delta}(J,\theta)=\frac{1}{2\pi\,D}\,\sum_{\vec m}\,e^{i(\gamma_{0}\,
m_{1}\,J-m_{2}\theta)}\,e^{-i\,1/2\,\gamma_{0} m_{1} m_{2}}\, 
~\hat{\cal E}_{N}^{m_{1}}\,\hat{\cal E}_{\phi}^{m_{2}}~, 
\label{4.11}
\end{equation}
where we used the particular $1/(2\pi\,D)$ normalization to examine the 
action-angle Wigner function   
and $J, \theta$ are introduced as the generalised 
{\it action-angle variables} as a physical realisation of the phase space 
vector ${\vec V} \to (\theta/\gamma_{0},I)$ in Eq's\,(\ref{3.1}). The 
change in the normalization factor from Eq.\,(\ref{3.1}) to Eq.\,(\ref{4.11}) 
is then simply the Jacobian of the transformation  
$d{\vec V} \to dI\,d\theta$.
The Wigner-Kirkwood basis $\hat{\Delta}(J,\theta)$ 
has the cyclic property that $\hat{\Delta}(J,\theta)=
\hat{\Delta}(J(mod D),\theta (mod 2\pi))$. 
Let's now insert the identity operator in (\ref{4.4})  
on both sides of the basis operators in (\ref{4.11}). Using 
Eq's\,(\ref{4.3}) 
and (\ref{4.5}) repeatedly $m_2$ and $m_1$ times, Eq.\,(\ref{4.11})  
becomes
\begin{equation}  
\hat{\Delta}(J,\theta)=\frac{1}{2\pi\,D}\,\sum_{\vec m}\,
\sum_{\ell=0}^{D-1}\,e^{i(\gamma_{0}\,m_{1}\,J-m_{2}\theta)}\,
e^{i\gamma_{0} \ell m_2}\,e^{i\gamma_{0} m_1 m_2/2}\,
\vert \phi\rangle_{\ell}\,~\,_{\ell+m_1}\langle \phi \vert\,~. 
\label{4.12}
\end{equation}
The action-angle Wigner function in any particular finite 
dimensional Hilbert space state $\vert \psi \rangle$ is then given as in 
(\ref{3.2}) by 
\begin{equation}
W_{\psi}(J,\theta)=\langle \psi \vert \,
\Delta(J,\theta)\,\vert \psi \rangle~, 
\label{4.13}
\end{equation}
with all required conditions for the generalised Wigner function  
satisfied. In Sec.VI.C we will examine the continuous limit of 
Eq.\,(\ref{4.13}) as $D \to \infty$.     

\section{The Limit to Continuum}
The large $D$ limit of the sine algebra has been extensively studied 
initially, for instance in Ref.s\,[5,6], and later by many other 
workers. We will not present these results here. We will also consider the 
$D \to \infty$ limit with the condition that $D$ remains a prime 
number. 

\subsection{The Number-Phase Basis}
In the limit $D \to \infty$ the spectra of $\hat{\cal U}$ and 
$\hat{\cal V}$ become arbitrarily dense and approach a 
continuously uniform distribution on the unit circle. Hence, for both  
unitary operators, the strong convergence is clearly guaranteed from 
those with discrete spectra to those with continuous spectra.\cite{25,26} 
In particular, the 
continuous limits of $\hat{\cal E}_{\phi}$ and $\hat{\cal E}_{N}$ will 
be identified as 
\begin{equation}
\begin{array}{rll}
\lim_{D \to \infty}\,\hat{\cal E}_{N}^{m_1} \to &
\hat{\tilde{\cal E}}_{N}^{\gamma} \equiv 
e^{-i\gamma\,\hat{N}}~, \qquad {\rm where} \qquad
&\gamma \equiv \lim_{D \to \infty} \, \frac{2\pi m_1}{D}
\in \mbox{\ee R} \\
\lim_{D \to \infty}\,\hat{\cal E}_{\phi}^{m_2} \equiv &
\hat{\tilde{\cal E}}_{\phi}^{m_2}
\qquad & 0 \le m_{2} < \infty~, \quad m_{2} \in \mbox{\ee Z}~. 
\end{array}
\label{5.1}
\end{equation}
where $\hat{\tilde{\cal E}}_{N}$ and $\hat{\tilde{\cal E}}_{\phi}$ 
are now corresponding unitary operators with continuous spectra. 
On the other hand, 
in the limit to continuity we must restrict the physical states that 
$\hat{\cal E}_{\phi}$ and $\hat{\cal E}_{N}$ act upon to 
those everywhere differentiable and continuous functions in the infinite 
dimensional Hilbert space. For all such acceptable states 
$\vert \psi\rangle$, the condition for weak convergence  
\begin{equation}
\lim_{D \to \infty}\,\Vert (\hat{\cal E}_{N}^{m_1}-
\hat{\tilde{\cal E}}_{N}^{\gamma}) \vert \psi \rangle \Vert^{2} < \epsilon 
\qquad {\rm where} \qquad \epsilon < 0^{+}~~{\rm (arbitrarily~ small)}, 
\label{5.2}
\end{equation}
and similarly for $\hat{\tilde{\cal E}}_{\phi}$, must be respected. 
In particular, it was shown in Ref.\,[25] that the eigenstates of 
$\hat{\tilde{\cal E}}_{N}$ and $\hat{\tilde{\cal E}}_{\phi}$ are   
good examples of such $\vert \psi\rangle$ and the convergence in (\ref{5.2}) 
in the limit $D \to \infty$ is known to exist. Considering the  
$D \to \infty$ limit of Eq.\,(\ref{4.3}) and (\ref{4.5}), the eigenstates 
of $\hat{\tilde{\cal E}}_{N}$ and $\hat{\tilde{\cal E}}_{\phi}$ are 
\begin{equation}
\vert n\rangle=\int\,\frac{d\phi}{\sqrt{2\pi}}\,e^{-i\phi\,n}\,
\vert \phi\rangle~, \qquad \vert \phi\rangle=\frac{1}{\sqrt{2\pi}}\,
\lim_{D \to \infty}\,
\sum_{n=0}^{D-1}\,e^{i\phi\,n}\,\vert n\rangle~, 
\label{5.3}
\end{equation}
where we have defined
\begin{equation}
\begin{array}{rl}
\lim_{D \to \infty}\,\vert n \rangle=&\vert n \rangle~, \qquad
0 \le n < \infty \\
\lim_{D \to \infty}\,\frac{1}{\sqrt{\gamma_{0}}}\,\vert \phi \rangle_{\ell}
\equiv& \vert \phi \rangle~,
\qquad \phi=\lim_{D \to \infty} \frac{2\pi\,\ell}{D}
\in \mbox{\ee R}~,\quad {\rm and} \quad 0 \le \phi < 2\pi~,
\end{array}
\label{5.4}
\end{equation}
with the proper normalizations $\langle \phi^{\prime}
\vert \phi\rangle=\delta(\phi-\phi^{\prime})$ and 
$\langle n^{\prime} \vert n \rangle=\delta_{n^{\prime},n}$. 
Remember that 
the periodic boundary conditions are still valid in the
 limit (i.e. $\vert \phi \rangle \equiv \vert
\phi+2\pi\rangle$
 and $\vert n \rangle=\lim_{D \to \infty}\,\vert n+D\rangle$).
For a generally acceptable state $\vert \psi\rangle=
\sum_{\ell=0}^{D-1}\,\psi_{\ell}\,\vert \phi\rangle_{\ell}$ with 
$\Vert \vert \psi \rangle \Vert=1$, a similar weak convergence condition 
as in (\ref{5.2}) stated for the phase operator requires 
\begin{equation}
\lim_{D \to \infty}\,\Vert (\hat{\cal E}_{\phi}-
\hat{\tilde{\cal E}}_{\phi}) \psi \rangle \Vert^{2}=
\lim_{D \to \infty}\,\sum_{\ell=0}^{D-1}\,
\vert \psi_{\ell}(e^{i\gamma_{0}\ell}-e^{i\phi})\,\vert^{2} < \epsilon~. 
\label{5.41}
\end{equation}
Since $\vert \psi_{\ell}\vert \le 1$, and the convergence  
\begin{equation}
\lim_{D \to \infty}\,\Vert (\hat{\cal E}_{\phi}-
\hat{\tilde{\cal E}}_{\phi}) \vert \psi \rangle \Vert^{2}=
\lim_{D \to \infty}\,
sup\Bigl\{\vert\,(e^{i\gamma_{0}\ell}-e^{i\phi})\vert^{2}:\, 
0 \le \ell < (D-1) \Bigr\} < \epsilon  
\label{5.42}
\end{equation}
is guaranteed because of Eq's\,(\ref{5.3}) and (\ref{5.4}),   
the only condition for the existence for such acceptable states is that
in the limit $D \to \infty$, the wavefunction $\psi_{\ell}$ is sufficiently 
well behaved and everywhere differentiable. Once the weak 
convergence condition in Eq.\,(\ref{5.2}) is satisfied for an acceptable 
state $\vert \psi \rangle$ expressed in one basis  
(i.e. in $\vert  n\rangle$ or $\vert \phi\rangle$), the weak convergence 
in the other basis is guaranteed by the Eq's\,(\ref{5.3}).   

The actions of the operators in (\ref{5.1}) on the infinite dimensional 
Hilbert space spanned by the vectors in (\ref{5.3}) are therefore   
\begin{equation}
\begin{array}{rlc}
\hat{\tilde{\cal E}}^{\gamma}_{N}\,\vert \phi \rangle=
\vert \phi-\gamma \rangle 
~, \qquad \hat{\tilde{\cal E}}^{\gamma}_{N}\,\vert n \rangle=&
e^{-i\gamma\,n}\,\vert n \rangle \\
\hat{\tilde{\cal E}}_{\phi}^{\ell}\,\vert n \rangle=\vert n-\ell \rangle~, 
\qquad 
\hat{\tilde{\cal E}}_{\phi}^{\ell}\,\vert \phi \rangle=&
e^{i\,\ell\,\phi}\,\vert \phi \rangle~.
\end{array}
\label{5.5}
\end{equation}
In this continuous limit, Eq.\,(\ref{4.7}) implies that
\begin{equation}
e^{-i\,\gamma\,\hat{N}}\,\hat{\tilde{\cal E}}_{\phi}^{\ell}=
e^{i\,\ell\,\gamma}\, 
\hat{\tilde{\cal E}}_{\phi}^{\ell}\,e^{-i\,\gamma\,\hat{N}}~. 
\label{5.6}
\end{equation}
Differentiating (\ref{5.6}) with respect to $\gamma$ and considering the 
limit $\gamma \to 0$ we find that   
\begin{equation}
\bigl[\hat{N},\hat{\tilde{\cal E}}_{\phi}^{\ell}]=
-\ell\,\hat{\tilde{\cal E}}_{\phi}^{\ell}
\label{5.7}
\end{equation}
which is the Susskind-Glogower-Carruthers-Nieto phase-number 
commutation relation\cite{27} with $\hat{\cal E}_{\phi}$ describing the 
unitary phase operator with a continuous spectrum as given in (\ref{5.5}). 
The expansion of Eq.\,(\ref{5.6}) for all orders in $\gamma$ is consistent 
with the first order term described in Eq.\,(\ref{5.7}). The coefficient 
of the $O(\gamma^{r})$ term reproduces the $r$th order commutation 
relations between $\hat{N}$ and $\hat{\tilde{\cal E}}_{\phi}^{\ell}$ as 
$[\hat{N},[\hat{N},\dots,[\hat{N},\hat{\tilde{\cal E}}_{\phi}^{\ell}
]\,] \dots ]=(-\ell)^{r}\,\hat{\tilde{\cal E}}_{\phi}^{\ell}$. In this 
respect, Eq.\,(\ref{5.6}) or, more generally, its discrete version in 
Eq.\,(\ref{1.6}) should be treated as generalized canonical commutation 
relations.   

\subsection{The spectrum shifted q-oscillator}
To study the $D \to \infty$ limit of the q-oscillator we first consider,    
in the numerator of $[n]$ in (\ref{4.8}), the equivalence of the sets  
of integers $\{n~{\vec m}\times {\vec m}^{\prime};
{\vec m}\times {\vec m}^{\prime} (mod D)\}_{0 \le n \le (D-1)}$
and $\{n\}_{0 \le n \le (D-1)}$ for any ${\vec m}, {\vec m}^{\prime}$. 
If ${\vec m}\times {\vec m}^{\prime} \ne 1$, this equivalence amounts to 
folding the value of $n{\vec m}\times {\vec m}^{\prime}$    
into the first {\it Brillouin zone} $n$ for $0\le n \le (D-1)$. In the  
limit, the spectrum is given by 
\begin{equation}
f(n)=\lim_{D \to \infty}\,\frac{1 \pm \sin(\gamma_{0}\,n)}{\vert 
\sin(\gamma_{0}{\vec m}\times {\vec m}^{\prime})\vert}~.  
\label{5.8}
\end{equation} 
Depending on ${\vec m}\times {\vec m}^{\prime}$, the sine term in the 
numerator takes continuous values in the range $[0,1)$ for 
 $0 \le n \le (D-1)$. Two limiting  
cases can be identified depending on the basis vectors  
${\vec m},{\vec m}^{\prime}$ by
\begin{equation}
f(n)=\cases{\lim_{D \to \infty} [1/\gamma_{0} \pm \, n], &if
${\vec m}\times {\vec m}^{\prime}=1$, \cr 
\lim_{D \to \infty} [1 \pm \gamma_{0}\,n], &if 
${\vec m}\times {\vec m}^{\prime}=(D-1)/4 \in \mbox{\ee Z}$~. \cr}
\label{5.9}
\end{equation}
The first case is identical to the continuous limit considered by 
Fujikawa\cite{18}. 
The spectrum is linear and unbounded, and the admissibility condition 
implies an unbounded positive shift by $\lim_{D \to \infty} 1/\gamma_0$. 
This is somewhat an infinitely shifted harmonic oscillator spectrum. 
Whereas, in the second case   
in (\ref{5.9}), one obtains a continuous, finite, and   
linear spectrum. The limit $D \to \infty$ has other interesting 
features. Fujikawa has shown that the vanishing of the index\cite{28}   
\begin{equation}
I=\sum_{n=0}^{D-1}\,\Bigl\{e^{-f(n)}\,-e^{-f(n+1)}\,\Bigr\}  
\label{5.10}
\end{equation}
is a stringent condition for the existence of the unitary phase operator.  
Using this index condition for the general admissible 
algebra in (54), it was previously shown\cite{19,28} that  
the limit $D \to \infty$ has a singular behaviour in the spectrum  
at $D=\infty$. This typical transition to a singular behaviour is also 
visible here if we compare the two indexes in (\ref{5.10}) once calculated  
using Eq.(\ref{5.8}) and then (\ref{5.9}). The former correctly 
yields $I=0$, whereas for the latter $I \ne 0$. Hence, in transition 
from (\ref{5.8}) to (\ref{5.9}), the vanishing index condition is violated. 
This proves that the spectrum as expressed in (\ref{5.9}) is not admissible 
at the limit $D=\infty$. The admissible form of (\ref{5.9}) is given by  
\begin{equation}
f(n)=\cases{\lim_{D \to \infty} 1/\gamma_{0}[1 \pm \,\sin(\gamma_{0}\,n)],  
&if ${\vec m}\times {\vec m}^{\prime}=1$, \cr
\lim_{D \to \infty} [1 \pm \sin(\gamma_{0}\,n)], &if
${\vec m}\times {\vec m}^{\prime}=(D-1)/4 \in \mbox{\ee Z}$ \cr}
\label{5.11}
\end{equation}
so that the vanishing index condition is respected. Thus, 
we learn that the vanishing index requires the information on the 
cyclic properties of the algebra to be maintained for all $D$ 
including the transition to infinity. For a more general consideration  
of the index theorem, we refer to Ref.\,[28]. Before closing  
this subsection, we mention as a side remark that the second limiting case 
in (\ref{5.11}) is somewhat similar to tight binding energy spectra in 
certain condensed matter systems. 

\subsection{The Wigner Function in the Phase Eigenbasis} 
Let's define in (\ref{4.11}) the variables 
$\phi=\lim_{D \to \infty}\,\phi_{\ell}~$~, 
$~~\phi+\gamma=\lim_{D \to \infty}\,\phi_{\ell+m_1}$ with  
$\phi,\gamma \in \mbox{\ee R}$ as well as 
$\vert \phi\rangle=\lim_{D \to \infty}\,\gamma_{0}^{-1/2}\,
\vert \phi\rangle_{\ell}$ in accordance with Eq's\,(\ref{5.1}) and 
(\ref{5.4}). Since  
$\phi,\gamma$ are continuous, we can replace the summation over $m_1$,   
in the limit, by an integral over $\gamma$ such that    
$\lim_{D \to \infty}\,1/D\,\sum_{\vec m}\, \to 
\lim_{D \to \infty}\,\sum_{m_2=0}^{D-1}\,\int\,d\gamma/2\pi$. 
Combining everything, we find for the Wigner function in this limit  
\begin{equation}
W_{\psi}(J,\theta)=\int\,\frac{d\gamma}{2\pi}\,
e^{i\,J\,\gamma}~~\langle \psi \vert \theta-\gamma/2\rangle\,
\langle \theta + \gamma/2 \vert \psi \rangle
\label{5.12}
\end{equation}
which is the conventional action-angle Wigner function represented 
in the continuous phase basis. Recently, a similar construction of the 
continuous Wigner function based on the continuous Weyl-Heisenberg basis 
was suggested n Ref.\,[13] as well as in 
Ref.\,[29] in very close correspondence with the 
results obtained here. Eq.\,(\ref{5.12}) can be realised as the action angle 
analog of Ref.\,[29]. If one starts in the generalised dual form represented 
by Eq.\,(\ref{3.1}) with the symmetric normalization, the discrete 
Weyl-Heisenberg representation of $\Delta({\vec V})$ is obtained which 
leads in the continuous limit to Wolf's Wigner function formulation  
in Ref.\,[29]. 
The continuous Weyl-Heisenberg representation as the standart 
representation of the Wigner function has also been 
examined by Schwinger\cite{4} as well as in Ref's.\,[8].  

\subsection{The continuously Shifted Finite Dimensional Fock Spaces and 
the Wigner Function in the generalised Fock Representation}
Let's consider the cyclic algebra in (\ref{4.1}) with the unitary phase and 
number operators as defined in Eq.s\,(\ref{4.3}) and (\ref{4.5}).    
We consider the phase operator $\hat{\cal E}_{\phi}^{-\alpha}$ in 
${\cal H}_{D}$ as   
\begin{equation}
\hat{\cal E}_{\phi}^{-\alpha}\,\vert \phi\rangle_{\ell}=
e^{-i\gamma_{0}\,\ell\,\alpha}\,\vert \phi\rangle_{\ell}~, \qquad {\rm and}
\qquad  
\hat{\cal E}_{\phi}^{-\alpha}\,\vert n \rangle \equiv 
\vert n + \alpha \rangle
\label{5.13}
\end{equation}
where $\alpha \in \mbox{\ee R}[0,1)$ and $\vert n + \alpha \rangle$ is 
defined by
\begin{equation}
\vert n + \alpha \rangle \equiv \frac{1}{\sqrt{D}}\,
\sum_{\ell=0}^{D-1}\,e^{-i\gamma_{0}\,
(n + \alpha)\ell}\,\vert \phi\rangle_{\ell}~.
\label{5.14}
\end{equation}
Since $\alpha \in \mbox{\ee R}[0,1)$, the states 
$\{\vert n+\alpha \rangle\}_{o \le n \le (D-1)}$ 
do not belong to the set of vectors spanning the finite dimensional 
conventional Fock space  
${\cal F}_{D}$. We now define a continuously shifted finite dimensional 
Fock space ${\cal F}_{D}^{(\alpha)}$ where 
$\{\vert n+ \alpha\rangle_{0 \le n \le (D-1)}; 
\alpha \in \mbox{\ee R}; \vert n +\alpha +D\rangle \equiv 
\vert n + \alpha \rangle \} 
\in {\cal F}_{D}^{(\alpha)}$. 
It can be readily verified that the following relations are 
satisfied by Eq.\,(\ref{5.14}) for all continuous values of $\alpha$:   
\begin{equation}
\langle n + \alpha \vert n^{\prime} + \alpha \rangle=\delta_{n,n^{\prime}}
~,\qquad 
\sum_{n=0}^{D-1}\,\vert n + \alpha \rangle\, \langle n + \alpha \vert=
\mbox{\ee I}~.   
\label{5.15}
\end{equation}
This implies that for a fixed $\alpha \in \mbox{\ee R}$, the shifted Fock 
space ${\cal F}_{D}^{(\alpha)}$ 
is also spanned by a complete orthonormal set of vectors 
$\{\vert n +\alpha \rangle\}_{0 \le n \le (D-1)}$ and it can equivalently  
be used in the generalised Fock representation of a physical state. The 
overlap between ${\cal F}_{D}$ and ${\cal F}_{D}^{(\alpha)}$ clearly 
respects the condition $\vert \langle n \vert n+\alpha\rangle \vert \le 1$ 
for all 
$\alpha \in \mbox{\ee R}$, and the extreme limits of 
$\alpha \to 0$ and $D \to \infty$ are commutative and well behaved: 
\begin{equation}
\vert \langle n \vert n+\alpha\rangle \vert=
\cases{\vert \sin{\pi\alpha}\vert/(\pi \alpha)~, & if ~~ 
$D \to \infty, ~~ 0 \le \alpha \le 1$ \cr 
1-(1-1/D)\,(\pi \alpha)^{2}/3!~, & if ~~ $D < \infty, ~~ 
\alpha \to 0 $\cr} 
\label{5.16}
\end{equation}
Since ${\cal F}_{D}$ and ${\cal F}_{D}^{(\alpha)}$ are spanned by cyclic 
vectors, $\alpha=1$ and $\alpha=0$ correspond to the identical Fock space 
representations.  
The action of the operator $\hat{\cal E}_{\phi}^{\beta}$ on the vectors in  
${\cal F}_{D}^{(\alpha)}$ is, therefore, equivalent to a continuous shift of 
the  
origin in ${\cal F}_{D}^{\alpha}$ by $\beta \in \mbox{\ee R}$ such that 
\begin{equation}
\hat{\cal E}_{\phi}^{(\beta)}\,:\, {\cal F}_{D}^{(\alpha)} \to  
{\cal F}_{D}^{(\alpha-\beta)}~.  
\label{5.17}
\end{equation}
Hence, a continuous shift $\beta$ induced by the operator 
$\hat{\cal E}_{\phi}^{(\beta)}$ is effectively equivalent to carrying  
vectors from the Fock space ${\cal F}_{D}^{(\alpha)}$ 
into the other one ${\cal F}_{D}^{(\alpha-\beta)}$, and the limit 
$\beta \to 0$ is continuous and analytic. Therefore, Eq.\,(\ref{5.17}) 
decribes an isomorphism between two inequivalent Fock spaces with equal 
dimensions. The physical    
implication of the state $\vert \alpha \rangle$ is that it corresponds 
to the {\it vacuum state} in ${\cal F}_{D}^{(\alpha)}$ and, unless 
$\alpha=0$, it is not the conventional vacuum $\vert 0\rangle$. The Fock 
space 
of the q-oscillator in Eq.\,(\ref{2.14}) is a typical example in which 
such a vacuum state is observed where we specifically have  
${\cal F}_{D}^{(D-1)/2}$. For $D$ being an odd integer,  
the conventional Fock representations in ${\cal F}_{D}$ are obtained. 
For $D$ being an even 
integer, the Fock space of the q-oscillator is  
${\cal F}_{D}^{1/2}$ and the vacuum state is $\vert 1/2\rangle$ 
correponding to $\alpha=1/2$. One crucial 
application of this is to examine the projection of the Wigner-Kirkwood 
basis onto the shifted Fock space ${\cal F}_{D}^{(\alpha)}$. Let's now 
insert the identity operator in (\ref{5.15}) on both sides of the unitary 
number-phase basis operators in (\ref{4.11}) yielding 
\begin{equation}
\hat{\Delta}(J,\theta)=\frac{1}{2\pi\,D}\,
\sum_{\vec m}\,e^{i(\gamma_{0}m_1 J-m_2 \theta)}\,e^{-i\gamma_{0}m_1 m_2/2}\,
\Bigl\{\sum_{n=0}^{D-1}\,\vert n+\alpha\rangle\langle n+\alpha\vert\Bigr\}\,
\hat{\cal E}_{\cal N}^{m_1}\hat{\cal E}_{\phi}^{m_2}\,
\Bigl\{\sum_{n^{\prime}=0}^{D-1}\,
\vert n^{\prime}+\alpha\rangle\langle n^{\prime}+\alpha\vert\Bigr\}~.  
\label{5.18}
\end{equation}
So far, the continuous shift $\alpha$ was arbitrary. Now, we adopt 
a particular set of values of $\alpha$ for each $J$ independently 
in such a way that 
$2\,(J-\alpha) \in \mbox{\ee Z}$. Since Eq.s\,(\ref{5.15}) are valid for  
all $\alpha \in \mbox{\ee R}$, this adaptive choice for $\alpha$ 
does not spoil the properties of  
the Wigner function studied in Sec.\,IV. Now, considering the limit 
$D \to \infty$ and following a similar calculation leading to (\ref{5.12}),   
we obtain the Wigner function 
\begin{equation}
W_{\psi}(J,\theta)=\frac{1}{2\pi}\,\lim_{D \to \infty}\,
\sum_{m_2=0}^{D-1}\,e^{-im_2 \theta}\,
\langle \psi \vert J-m_2/2\rangle\,\langle J+m_2/2\vert \psi \rangle  
\label{5.19}
\end{equation}
which is expressed in the 
shifted Fock bases in the limit $D \to \infty$.  
For the choice of $\alpha$ as  
$2\,(J-\alpha) \in \mbox{\ee Z}$, we have for the basis vectors 
$\{\vert J \pm m_2/2\rangle; m_2=odd\} \in {\cal F}_{D}^{(\alpha \pm 1/2)}$ 
and 
$\{\vert J \pm m_2/2\rangle; m_2=even\} \in {\cal F}_{D}^{(\alpha)}$.
Note that because of the cyclic property of the vectors in ${\cal H}_{D}$, 
the shifted Fock space ${\cal F}_{D}^{\alpha+1/2}$ shares the 
same vectors with ${\cal F}_{D}^{\alpha-1/2}$ for 
all $D < \infty$ and $\alpha \in \mbox{\ee R}[0,1)$. Thus, 
${\cal F}_{D}^{\alpha+1/2}$ and ${\cal F}_{D}^{\alpha-1/2}$ are indeed the 
same shifted Fock space. This discussion implies that if in the  
summation in (\ref{5.19}), the even and 
odd values of $m_2$ are separated, the Wigner function becomes a sum of two 
contributions $W^{(even)}$ and $W^{(odd)}$ projected onto 
${\cal F}_{D}^{\alpha}$ and ${\cal F}_{D}^{\alpha+1/2}$ for even  
and odd $m_2$ respectively as 
\begin{equation}
W_{\psi}(J,\theta)=W^{(even)}_{\psi}(J,\theta)+
W^{(odd)}_{\psi}(J,\theta)~.  
\label{5.20}
\end{equation}
Since each contribution is based on a different $D$ dimensional shifted 
Fock basis, they are properly normalised. 
It is interesting to note that a similar decomposition 
of the Wigner function in the Fock representation  
has been recently proposed by Luk\v{s} and Pe\v{r}inova\cite{30} as well 
as by Vaccaro\cite{24} in order to avoid certain superficial 
anomalies of the Wigner function they use in mixed physical states. 
Using continuously shifted Fock spaces, the decomposition they propose 
follows naturally.  
To elaborate more on the resolution of the anomalous behaviour of the 
Wigner function using the shifted Fock spaces exceeds our purpose here.  
It can be shown that the concept of continuously shifted Fock basis can 
also be generalised to the continuously shifted discrete Schwinger basis 
vectors $\{\,\vert u\rangle_{k}\,\}$ and $\{\,\vert v\rangle_{k}\,\}$. 
This subtle point certainly deserves much more attention in the generalised 
formulation of the Wigner function and quantum canonical transformations, 
which we intend to present in a forthcoming work. 

\section{Conclusions}
The central theme of this work was to demonstrate that 
conceptual foundation of the quantum phase lies in the algebraic 
properties of the canonical transformations on the generalised   
quantum phase space. In this context:  

1) It is shown that the Schwinger operator basis provides subalgebraic  
realisations of the admissible q-oscillators 
in addition to the known deformed $su(2)$ symmetries labeled 
by the lattice vectors in $\mbox{\ee Z}_{D}\times \mbox{\ee Z}_{D}$.   
The intensively studied magnetic translation operator algebra is a specific 
physical realisation of Schwinger's operator algebra. In this context, 
some interesting physics might be found in the realisation of the shifted 
q-oscillator subalgebra in terms of the magnetic translation operators 
as applied to the Bloch electron problem. To the author's knowledge, the 
nearest approach to 
this idea has been made by Fujikawa et al. recently [see the second 
reference in (18)].   

2) Certain equivalence classes within each subalgebra, using  
different lattice labels, are identified in terms of area preserving  
transformations. A general formulation of such discrete, linear canonical 
transformations is presented. 

3) The dual form between the Schwinger operator basis and the generalised 
discrete Wigner-Kirkwood basis is examined and the connection with the  
general area preserving canonical transformations on  
$\mbox{\ee Z}_{D}\times \mbox{\ee Z}_{D}$ is briefly studied. 

4) The application of the Schwinger operator basis on the number-phase basis 
is discussed and shown that it provides an algebraic approach to   
the formulation of the quantum phase problem. The admissibly shifted 
q-oscillator realisations of  
the Schwinger basis are studied from this algebraic point of view.
The generalised Wigner-Kirkwood basis is examined in the unitary   
number-phase basis and the limit to the conventional formulation of the 
action-angle Wigner function is 
investigated as the size of the lattice tends to infinity,  
or reciprocally, as the lattice spacing $2\pi/D$ tends to zero. 

5) Finally, much work has to be done on understanding the quantum phase 
problem within the canonical quantum phase space formalism. This problem is 
evidently connected 
with the recent research areas such as classical and quantum integrability, 
the deformation quantization, theory of nonlinear quantum canonical 
transformations and the Lie algebraic representations of the Wigner function. 
\acknowledgements{
The author is grateful to Professor T. Dereli (Middle East Technical 
University),  
Professor M. Ar{\i}k (Bo\u{g}azi\c{c}i University), and 
Professor A. Klyachko (Bilkent University) for useful discussions 
and critical comments.


\begin{thebibliography}{99}
\bibitem{1} X. Shen, Int. J. Mod. Phys. {\bf A7}, 3717 (1992). 
\bibitem{2} J. Zak, Phys. Rev. {\bf 134}, 1602 (1964); 
E. Brown, Phys. Rev. {\bf 133}, 1038 (1964); T. Dereli and A. Ver\c{c}in, 
Phys. Lett. {\bf B288}, 109 (1992); ibid, J. Phys. {\bf A26}, 6961 (1993). 
\bibitem{3} P.B. Wiegmann and A.V. Zabrodin, Phys. Rev. Lett. {\bf 72}, 1890 
(1994); Guang-Hong Chen, Le-Man Kuang and Mo-Lin Ge, Phys. Rev. 
{\bf B53}, 9540 (1996); ibid, Phys. Lett. {\bf A213}, 231 (1996). 
\bibitem{4} J. Schwinger, Proc. Nat. Acad. Sci. {\bf 46}, 883, 1401 (1960). 
\bibitem{5} E.G. Floratos, Phys. Lett. {\bf B228}, 335 (1989). 
\bibitem{6} D. Fairlie, P. Fletcher and C. Zachos, Phys. Lett. {\bf B218},
203 (1989); D.B. Falirlie and C. Zachos, Phys. Lett. {\bf B224}, 101 (1989).
\bibitem{7} V. Arnold, Ann. Inst. Fourier, XVI, no 1, 319 (1966); 
V. Arnold, {\it Mathematical Methods of Classical Mechanics} (Springer, 
Berlin, 1978). 
\bibitem{8} R. Aldrovandi and D. Galetti, J. Math. Phys. {\bf 31}, 2987
(1990).
\bibitem{9} I. I. Kogan, Int. J. Mod. Phys. {\bf A9}, 3887 (1994). 
\bibitem{10} Choon-Lin Ho, J. Phys. {\bf A29}, L107 (1996).
\bibitem{11} S. Ito, D. Karabali and B. Sakita, Phys. Lett. {\bf B296}, 
143 (1992); B. Sakita, Phys. Lett. {\bf B315}, 124 (1993).  
\bibitem{12} I.I. Kogan, Mod. Phys. Lett. {\bf A7}, 3717 (1992). 
\bibitem{13} D. Galetti and A.F.R. de Toledo Piza, Physica {\bf 149 A},
267 (1988).
\bibitem{14} E.P. Wigner, Phys. Rev. {\bf 40}, 749 (1932).
\bibitem{15} J.G. Kirkwood, Phys. Rev. {\bf 44}, 31 (1933). 
\bibitem{16} R. Balian and C. Itzykson, C.R. Acad. Sc. Paris, {\bf 303} 773
(1986).
\bibitem{17} H. Weyl, The Theory of Groups in Quantum Mechanics
(NY, Dover 1931).
\bibitem{18} K. Fujikawa, L.C. Kwek and C.H. Oh, Mod. Phys. Lett.
{\bf A 10}, 2543 (1995). 
\bibitem{19} T. Hakioglu, J. Phys. {\bf A 31}, 707 (1998); 
Hong-Chen Fu and Ryu Sasaki, J. Phys. {\bf A29}, 4049 (1996).
\bibitem{20} K. Fujikawa and Harunobu Kubo, Mod. Phys. Lett. {\bf A12}, 403 
(1997); K. Fujikawa and Harunobu Kubo, Phys. Lett. {\bf A239}, 21 (1998).  
\bibitem{21} M. Hillery, R.F. O'Connell, M.O. Scully and E.P. Wigner, 
Phys. Rep. {\bf 106}, 121 (1984). 
\bibitem{22} William K. Wooters, Ann. Phys. {\bf 176}, 1 (1987).
\bibitem{23} Jo\~ao P. Bizzaro, Phys. Rev {\bf A49}, 3255 (1994). 
\bibitem{24} John Vaccaro, Phys. Rev. {\bf A52}, 3474 (1995). 
\bibitem{25} E.C. Lerner, H.W. Huang and G.E. Walters, J. Math. Phys. 
{\bf 11}, 1679 (1970)
\bibitem{26} F. Riesz and B.Sz. Nagy, {\it Functional Analysis}, 
(Ungar, Newyork, 1955).
\bibitem{27} L. Susskind and J. Glogower, Physics {\bf 1}, 49 (1964);
P. Carruthers and M.M. Nieto, Phys. Rev. Lett. {\bf 14}, 387 (1965); ibid,
Rev. Mod. Phys. {\bf 40}, 411 (1965).
\bibitem{28} K. Fujikawa, Phys. Rev. {\bf A52}, 3299 (1995).  
\bibitem{29} K.B. Wolf, Opt. Commun. {\bf 132}, 343 (1996). 
\bibitem{30} A. Luk\v{s} and V. Pe\v{r}inov\'a, Phys. Scr. {\bf T48}, 94 
(1993).
\end{thebibliography}
\end{document}